\documentclass[journal]{IEEEtran} 

\ifCLASSINFOpdf

\else

\fi

\usepackage{subfigure}
\usepackage{graphicx}
\usepackage{amssymb}
\usepackage{amsmath}
\usepackage{amsfonts}
\usepackage{epsfig}
\usepackage{epstopdf}
\usepackage{xr}
\usepackage{float}
\usepackage{xspace}
\usepackage{algorithm}
\usepackage{algpseudocode}
\usepackage{algorithmicx}
\usepackage{etex}
\usepackage{multirow}
\usepackage{color}

\begin{document}

\title{
Optimization Algorithms for Catching Data Manipulators in Power System Estimation Loops
}

\author{Mang~Liao
           and~Aranya~Chakrabortty,~\IEEEmembership{Senior Member,~IEEE}
\thanks{Mang Liao and Aranya Chakrabortty are with the Department of Electrical and Computer Engineering,
        North Carolina State University, Raleigh, NC 27695, USA
        {\tt\small mliao@ncsu.edu, achakra2@ncsu.edu} }
\thanks{This work is supported by NSF grant ECCS 1544871. }}

\maketitle

\begin{abstract}
In this paper we develop a set of algorithms that can
detect the identities of malicious data-manipulators in distributed
optimization loops for estimating oscillation modes in large power
system models. The estimation is posed in terms of a consensus problem among multiple local estimators that jointly solve for the characteristic polynomial of the network model. If any of these local estimates are compromised by a malicious attacker, resulting in an incorrect value of the consensus variable, then the entire estimation loop can be destabilized. We present four iterative algorithms by which this instability can be quickly detected, and the identities of the compromised estimators can be revealed. The algorithms are solely based on the computed values of the estimates, and do not need any information about the model of the power system.  Both large and covert attacks are considered. Results are illustrated using simulations of a IEEE 68-bus power system model.

\end{abstract}


\section{Introduction}

The wide-area measurement systems (WAMS) technology using Phasor Measurement Units (PMUs) has been regarded as the key to guaranteeing stability, reliability, state estimation, control, and protection of next-generation power systems \cite{phadke}. However, with the exponentially increasing number of PMUs deployed in the North American grid, and the resulting explosion in data volume, the design and deployment of an efficient wide-area communication and computing infrastructure, especially from the point of view of resilience against a wide varieties of nefarious cyber-physical attacks, is evolving as one of the greatest challenges to the power system and IT communities.
With several thousands of networked PMUs being scheduled to be installed in the United States by 2020, exchange of Synchrophasor data between balancing authorities for any type of wide-area control will involve several thousands of Terabytes of data flow in real-time per event, thereby opening up a wide spectrum of opportunities for adversaries to induce data manipulation attacks, denial-of-service attacks, GPS spoofing, attacks on transmission assets, and so on. The challenge is even more aggravated by the gradual transition of WAMS from centralized to distributed in order to facilitate the speed of data processing~\cite{behzad2} and \cite{Jianhua}.

Several recent papers such as \cite{Liu2011False}-\cite{Sridhar2012} have studied how false-data injection attacks may be deceptively injected into a power grid using its state estimation loops, and through data in relays and intelligent electronic devices (IEDs) \cite{Ten2011}.
Others have proposed estimation-based mitigation strategies to secure the grid against many of these attacks  \cite{Vukovic2012Network}-\cite{deepa}. The fundamental approach behind many of these designs is based on the so-called idea of Byzantine consensus  \cite{Tseng2015Fault}-\cite{sundaram}, a fairly popular topic in distributed computing, where the goal is to drive an optimization or optimal control problem to a near-optimal solution despite the presence of a malicious agent. In practice, however, this approach is not acceptable to most WAMS operators as they are far more interested in finding out the {\it identity} of a malicious agent if it exists in the system, disconnect it from the estimation or control loop, and continue operation using the remaining non-malicious agents rather than settling for a solution that keeps the attacker unidentified in the loop. This basic question of how to {\it catch} malicious agents in distributed wide-area monitoring applications is still an open challenge in the WAMS literature.

In this paper we address this problem in the context of identifying malicious data-manipulators in distributed optimization loops for wide-area oscillation monitoring. The specific application of our interest is the estimation of electro-mechanical oscillation modes or eigenvalues from streaming PMU data following a small-signal disturbance in the grid \cite{mani}. Accurate estimation of these oscillation modes is critical to grid operators, as many control and protection decisions depend on whether the damping and residues of the modes are below a chosen threshold or not. The fundamental set-up for the distributed optimization is based on the Alternating Direction Multiplier Method (ADMM) \cite{boyd}. The power system is divided into multiple non-overlapping areas, each equipped with a local estimator. These local estimators use local sensor measurements of voltage, phase angle, and frequency from PMUs to carry out a local regression algorithm for generating a local estimate of the characteristic polynomial of the system, and, thereafter, communicate this estimate to a central supervisor. The supervisor computes the average or {\it consensus} of all estimates, and broadcasts this consensus variable back to each local estimator to be used in the next round of regression. If all estimates are accurate, then ADMM is guaranteed to converge asymptotically to the true optimal solution of the characteristic polynomial, following which the central supervisor can solve for its roots to obtain the desired eigenvalues. However, if one or more of the local estimates are manipulated by attackers, then the resulting consensus variable will be inaccurate, which, in turn, will contaminate the accuracy of every local estimate. Even a small amount of bias at a single iteration can destabilize the entire estimation process. To combat this, we first develop an algorithm to show how ADMM can be used by the central supervisor to catch the identities of malicious estimators by simply tracking the quality of every incoming estimate. This algorithm, however, can become computationally expensive if the network size is large. Therefore, we propose another algorithm where the central supervisor, instead of computing the average, employs a Round-Robin technique to generate the consensus variable, and show that by tracking the evolution of only this consensus variable it is possible to identify the malicious estimators. Both large and covert attacks are considered. We analyze the convergence properties of the proposed algorithms, and illustrate their effectiveness using simulation results on a IEEE 68-bus power system model.

Preliminary results on this problem for single-node attacks have been reported in our recent paper \cite{LiaoA16}. This paper expands those results to a more general case where attacks may be imposed on multiple nodes, and the attack magnitude can be either large or small, leading to a much more complex set of algorithms and simulations. Moreover, unlike \cite{LiaoA16}, here we also show how the standard ADMM itself can be used for attack detection for both large and covert attacks.

The remaining sections of the paper are organized as follows. Section \uppercase\expandafter{\romannumeral2} presents the power system model and the mode estimation problem of interest, and formulates the attack localization problem. Section \uppercase\expandafter{\romannumeral3} shows how to use standard ADMM (S-ADMM) and its Round-Robin (RR) version for localization of biases with noticeable magnitudes, respectively. Section \uppercase\expandafter{\romannumeral4} extends the idea to identify malicious users with multiple small biases. Section \uppercase\expandafter{\romannumeral5}
compares the RR-ADMM with S-ADMM in terms of their convergence properties. Section \uppercase\expandafter{\romannumeral6} illustrates all the algorithms through simulations of a IEEE 68-bus power system model. Section \uppercase\expandafter{\romannumeral7} concludes the paper.

\section{ Mode Estimation using Distributed ADMM}\label{section2}

\subsection{Problem Set up Using Standard ADMM}

Consider a power system with $n$ synchronous generators  and $n_l$ loads connected by a given topology. Each synchronous generator is modeled by a second-order swing equation, and each bus is modeled by two algebraic equations for active and reactive power balance. The measured output ${\boldsymbol {y}}(t) \in \mathcal R^p$ is considered to be a set of  measurements of small-signal values of bus voltages, phase angles, and frequencies, denoted, in general, by $y_i(t)$, $i=1,\ldots,p$, measured at $p \leq (n+n_l)$ designated buses using PMUs. Following a disturbance, which can be simply assumed to be a unit impulse function, the impulse response of the $i^{th}$ output can be written as
\begin{align}
y_{i}(t)=\sum_{k=1}^{2n} r_{i,k}e^{(-\sigma_k+j\Omega_k) t}+r^*_{i,k}e^{(-\sigma_k-j\Omega_k) t},
\label{exp_form}
\end{align}
for $i=1, \ldots, p$, where $\lambda_k=(-\sigma_k \pm j\Omega_k),~(j=\sqrt{-1})$, are the complex eigenvalues (or modes) of the small-signal electro-mechanical model of the power system. $\Omega_k>0$ and $\sigma_k>0$ are the frequency and damping factor of the $k^{th}$ mode, respectively, while $r_{ik}$ is the residue of the $k^{th}$ mode in the $i^{th}$ output. In order to estimate these $2n$ complex eigenvalues $\lambda_k$ from ${\boldsymbol {y}}(t)$ one may employ centralized techniques such as Prony's algorithm. We briefly recapitulate that algorithm to motivate our problem. Sampling $y_i (t)$ with a uniform sampling period of $T$, a generic expression for the $z$-transform of $y_i(m)\triangleq y_{i}(t)|_{t=mT}$, ($m=0,1,\ldots,M$), can be written as
\begin{align}
y_i(z)=\frac{b_{0i}+b_{1i}z^{-1}+b_{2i}z^{-2}+\cdots+b_{2ni}z^{-2n}}{1+a_1z^{-1}+a_2z^{-2}+\cdots+a_{2n}z^{-2n}},
\label{dis}
\end{align}
where $a$'s and $b$'s are constant coefficients of the characteristic polynomial and the zero polynomial, respectively. The roots of the characteristic polynomial will provide the discrete-time poles of the system. One can, therefore, first estimate the coefficient vector ${\boldsymbol{a}}:\{a_1,\dots,a_{2n}\}$, compute the discrete-time poles, and finally convert them to the continuous-time poles to obtain $\sigma_k$ and $\Omega_k$, for $k=1,\ldots,2n$,  as follows \cite{mani}:

{\it Step 1.} Solve for $\boldsymbol{a}$ from
\begin{align}
\hspace{-0.1em} \underset{{\boldsymbol{c}}_i}{\underbrace{\begin{bmatrix} y_i(2n) \\ y_i(2n+1) \\ \vdots \\ y_i(2n+\ell) \end{bmatrix}}}=\underset{{\boldsymbol H}_i}{\underbrace{\begin{bmatrix}
y_i(2n-1) & \cdots & y_i(0) \\
y_i(2n) & \cdots & y_i(1)\\
\vdots &  & \vdots \\
y_i(2n+\ell-1) &\cdots & y_i(\ell)
\end{bmatrix}}}~~\underset{{\boldsymbol{a}}}{\underbrace{\begin{bmatrix}
-a_1\\-a_2\\ \vdots \\ -a_{2n}
\end{bmatrix}}}, \label{prony1}
\end{align}
where $\ell$ is an integer satisfying $2n+\ell \leq M-1$. Concatenating ${\boldsymbol{c}}_i$ and ${\boldsymbol H}_i$ in (\ref{prony1}) for $i=1,\ldots,p$, one can find $\boldsymbol{a}$ by solving a least-squares (LS) problem
\begin{align}
\underset{\boldsymbol{a}}{\mathrm{min}}~\frac{1}{2} ||\begin{bmatrix} {\boldsymbol H}_1 \\ \vdots \\ {\boldsymbol H}_p \end{bmatrix} {\boldsymbol{a}}-\begin{bmatrix} {\boldsymbol{c}}_1 \\ \vdots \\ {\boldsymbol{c}}_p \end{bmatrix}||^2,
\label{31}
\end{align}
where $|| \cdot ||$ denotes the 2-norm of a vector.

{\it Step 2.} Find the roots of the discrete-time characteristic polynomial, say denoted by ${\boldsymbol z}_k$, $k=1,\ldots,2n$. Then, the desired eigenvalues $\lambda_k$ are equal to $\ln({\boldsymbol z}_k)/T$.

\begin{figure}[!t]
\centering
\includegraphics[width=8.5cm,height=6.5cm]{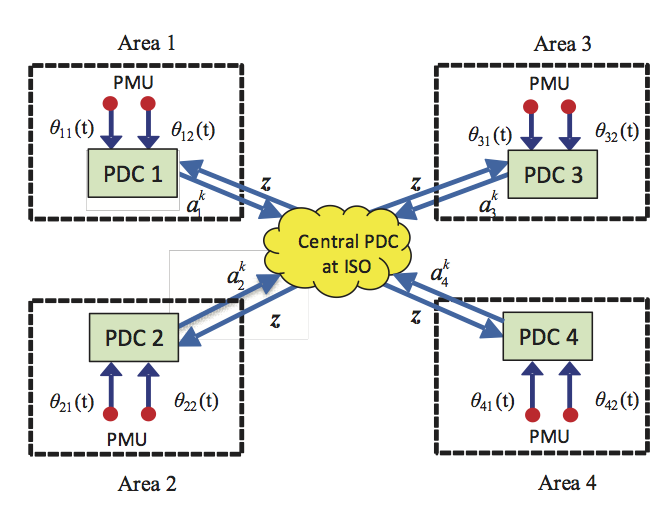}
\caption{Distributed architecture for a 4-area power system network }
\label{fig1}
\end{figure}

The centralized approach, however, becomes computationally untenable as more and more PMUs are installed in the system. Instead a distributed solution is much more preferable. Fortunately, the LS problem (\ref{31}) can be reformulated as a global consensus problem over a distributed network of $N$ computational areas spanning the entire grid, as shown in our recent paper \cite{behzad2}. We briefly recall that distributed architecture before stating our main problem statement. An example of the architecture with four areas is shown in  Fig. \ref{fig1}. We assume each area to be equipped with one local  Phasor Data Concentrator (PDC) (located at its control center). These PDCs receive local PMU measurements, run a local LS using these measurements, and then share the estimated parameters with a supervisory PDC at the ISO. For convenience, we will refer to the PDCs inside the areas as `local PDC' and the PDC at the ISO as `central PDC' as indicated in Fig. \ref{fig1}. Let the indices of the PMUs in the $m^{th}$ area be $i_1, i_2, \ldots, i_{m_i}$. The problem (\ref{31}) then can be rewritten as
\begin{align}\label{p1}
\min_{{\boldsymbol{a}}_1,\ldots,{\boldsymbol{a}}_N,\mathbf{z}} \sum_{i=1}^N \frac{1}{2} ||\hat{{\boldsymbol{H}}}_i{\boldsymbol{a}}_i-\hat{\boldsymbol{c}}_i||^2, \quad \mbox{s.t}\quad {\boldsymbol{a}}_i-{\boldsymbol{z}}=0,
\end{align}
for $i=1,\ldots,N$, where ${\boldsymbol{a}}_i$ is the vector of the \emph{primal variables}, $\boldsymbol{z}$ is the \emph{global consensus variable}, $\hat{\boldsymbol{H}}_i=[{\boldsymbol H}_{i_1}^{\rm T}, {\boldsymbol H}_{i_2}^{\rm T}, \ldots, {\boldsymbol H}_{i_{m_i}}^{\rm T}]^{\rm T}$, and $\hat{\boldsymbol{c}}_i=[{\boldsymbol c}_{i_1}, {\boldsymbol c}_{i_2}, \ldots, {\boldsymbol c}_{i_{m_i}}]^{\rm T}$. Each block element of ${\hat{\boldsymbol H}}_i$ and ${\hat{\boldsymbol c}}_i$ can be constructed after the disturbance using the data matrices shown in \eqref{prony1}. The estimators can wait up to a certain number of samples, say $2n+\ell$ as indicated in \eqref{prony1}, and gather the local measurements up to that iteration. The iteration index $k=0$ in the algorithms stated hereafter starts from this instant of time. The global consensus solution of (\ref{p1}) is achieved when the local estimations of the $N$ regional PDCs, ${\boldsymbol{a}}_i$, $\forall~i=1,\ldots,N$, reach the same value.  To solve (\ref{p1}) in a distributed way, we use ADMM \cite{boyd}, which reduces to the following set of recursive updates by using an \emph{augmented Lagrangian} (for details, please refer to \cite{behzad2}):
\begin{subequations}
\label{p3}
\begin{align}
&{\boldsymbol{w}}_i^{k}={\boldsymbol{w}}_i^{k-1}+\rho({\boldsymbol{a}}_i^{k}-{\boldsymbol{z}}^{k}), \label{p3_1}\\
&{\boldsymbol{a}}_i^{k+1}=((\hat{\boldsymbol{H}}_i)^T\hat{\boldsymbol{H}}_i+\rho {\boldsymbol{I}})^{-1}((\hat{\boldsymbol{H}}_i)^T\hat{\boldsymbol{c}}_i-{\boldsymbol{w}}^{k}_i+\rho {\boldsymbol{z}}^{k}), \label{p3_2}\\
&{\boldsymbol{z}}^{k+1}=\frac{1}{N}\sum_{i=1}^{N}{\boldsymbol{a}}_{i}^{k+1} , \label{p3_3}
\end{align}
\end{subequations}
where ${\boldsymbol{w}}_i$ is the vector of the \emph{dual variables}, or the Lagrange multipliers associated with (\ref{p1}), and $\rho>0$ denotes the \emph{penalty factor}. To distinguish it from another variant of ADMM to be proposed later in the paper, we will refer to (\ref{p3}) as the standard ADMM, or S-ADMM in short. In \cite{behzad2}
we developed the cyber-physical architecture by which local PDCs and the central PDC can exchange information between each other for executing S-ADMM. We summarize that architecture as follows. Consider the $k^{th}$ iteration. 
 \textbf{Step~1)} any local PDC $i$ runs the dual-primal update for (${\boldsymbol{w}}_i^{k}$, ${\boldsymbol{a}}_i^{k+1}$) using (\ref{p3_1}) and (\ref{p3_2}), after receiving the consensus variable ${\boldsymbol{z}}^{k}$ from the central PDC; \textbf{Step~2)} the local PDC $i$ transmits ${\boldsymbol{a}}_i^{k+1}$ to a central PDC; \textbf{Step~3)} the central PDC calculates the consensus variable  ${\boldsymbol{z}}^{k+1}$ using (\ref{p3_3}); \textbf{Step~4)} the central PDC broadcasts ${\boldsymbol{z}}^{k+1}$ to the local PDCs in each area for their next update. Since the LS problem is convex, therefore as $k \rightarrow \infty$, ${\boldsymbol z}^k$ in \eqref{p3_3} converges to ${\boldsymbol z}^*$ which is the solution of the centralized problem \eqref{p1}. Also, because of  consensus, every ${\boldsymbol a}_j^k$ converges to ${\boldsymbol z}^*$, $1 \le j \le N$; \textbf{Step~5)} finally, the central PDC estimates the eigenvalues of the small-signal model by solving for the roots of the  characteristic polynomial given by ${\boldsymbol z}^*$. Detailed discussions on the scalability and convergence for this ADMM-based mode estimation problem can be found in \cite{behzad2} and \cite{Jianhua}. We skip those results here for the sake of brevity.


\subsection{Problem Formulation for Attack Identification} \label{sec2}
Since companies may not employ expensive dedicated fiber optic networks for message passing with the ISO, the communication between the local PDCs and the central PDC is most likely to happen over an open wide-area communication network, making it is easy for attackers to hack into the PDCs, and corrupt the value of their estimates. The attack model is defined as follows. The attacker has the knowledge of the data matrices ${\hat{\boldsymbol H}}_i$ and ${\hat{\boldsymbol c}}_i$, and the estimates ${\boldsymbol a}_i^k$ and ${\boldsymbol w}_i^{k-1}$ $\forall k$, with $i \in \mathcal{S}$ where $\mathcal{S}$ is the set of indices of all attacked PDCs.  
Because the ISO does not know that the message is corrupted, it will still calculate the consensus variable by averaging the estimates obtained from all individual local PDCs. Thus ${\boldsymbol z}^k$ in \eqref{p3_3} will become
\begin{align}\label{z_bias}
{\boldsymbol z}^k = \Delta^k+ \frac{1}{N}\left(\sum_{i=1}^N{\boldsymbol a}_i^k\right),
\end{align}
where $\Delta^k = \frac{1}{N}\left(\sum\limits_{j = 1,j \in \mathcal{S}} \Delta_j^k\right)$. Notice that here we consider any number of local PDCs to be attacked, as long as there is at least one unattacked PDC, and that the bias $\Delta_j^k$ may be time-varying and of arbitrary magnitude.
Denoting $A_j := (({\hat {\boldsymbol H}}_j)^{\rm T}{\hat {\boldsymbol H}}_j + \rho {\boldsymbol I}_{2n})^{-1}$, and $C_j := ({\hat {\boldsymbol H}}_j)^{\rm T}{\boldsymbol c}_j$, following the expression of the consensus variable with bias as in \eqref{z_bias}, the S-ADMM algorithm in \eqref{p3} can be written in a state-variable form:
\begin{align}\label{Wang_model}
\left[ {\begin{array}{*{20}{c}}
{{{\boldsymbol a}^{k + 1}}}\\
{{{\boldsymbol a}^{k }}}
\end{array}} \right] = \underbrace {\left[ {\begin{array}{*{20}{c}}
{{{\boldsymbol L}_{11}}}&{{{\boldsymbol L}_{12}}}\\
{{{\boldsymbol I}}}&{{{\boldsymbol 0}}}
\end{array}} \right]}_{\boldsymbol L}\left[ {\begin{array}{*{20}{c}}
{{{\boldsymbol a}^k}}\\
{{{\boldsymbol a}^{k-1}}}
\end{array}} \right] + \left[ {\begin{array}{*{20}{c}}
{{{\boldsymbol P}}}\\
{{{\boldsymbol 0}}}
\end{array}} \right]\Delta^k,
\end{align}
where ${\boldsymbol P} = \left[ {\begin{array}{*{20}{c}}
{\rho A_1}\\
{\rho A_2}\\
{\vdots}\\
{\rho A_N}
\end{array}} \right] $, ${{\boldsymbol L}_{12}} = -\left[ {\begin{array}{*{20}{c}}
{\frac{{\rho {A_1}}}{N}}& \ldots &{\frac{{\rho {A_1}}}{N}}\\
 \vdots & \ddots & \vdots \\
{\frac{{\rho {A_N}}}{N}}& \ldots &{\frac{{\rho {A_N}}}{N}}
\end{array}} \right]$, and
\begin{align}
&{{\boldsymbol L}_{11}} = \nonumber\\
&\left[ {\begin{array}{*{20}{c}}
{I + \frac{{(2-N )\rho }}{N}{A_1}}&{\frac{2\rho A_1}{N}}& \ldots &{\frac{2\rho A_1}{N}}\\
{\frac{2\rho A_2}{N}}&{I + \frac{{(2-N )\rho }}{N}{A_2}}&\ldots &{\frac{2\rho A_2}{N}}\\
 \vdots & \vdots  & \vdots&\vdots \\
{\frac{2\rho A_N}{N}} &\ldots& {\frac{2\rho A_N}{N}}&{I + \frac{{(2-N )\rho }}{N}{A_N}}
\end{array}} \right].
\end{align}

The rows of $L$ in \eqref{Wang_model} add up to 1, and so when $\Delta_j^k=0$ then the trajectories of ${\boldsymbol a}_i^k$ for every $i=1,\ldots,N$ converge asymptotically to consensus, as shown in \cite{boyd}. However, when the arbitrary disturbance $\Delta_j^k$ is added then these trajectories will diverge unless $\Delta_j^k$ is chosen in a special way so that its entries corresponding to the consensus modes are exactly zeros. That, however, is very unlikely to happen as the attacker will not know the matrix $L$ prior to the attack, and, therefore, cannot use any information about its consensus properties for designing $\Delta_j^k$. In any case, the attacker would benefit most if ${\boldsymbol a}_i^k$ start diverging, implying that she has been able to destabilize the estimation loop. Detecting the identity of the corrupted PDC is, therefore, is crucial to retain normal operation of the loop. In the following sections, we propose a variety of algorithms to catch the identity of these data manipulators, starting with S-ADMM and then its round-robbin version. We also consider the case when $\Delta_k^j$ is small or covert, and show how S-ADMM can be used for the detection by reducing the penalty parameter $\rho$. In that situation the round-robbin algorithm can also detect the identity of the manipulators by monitoring the dual variable ${\boldsymbol w}_j^k$ without requiring any knowledge of the individual estimates ${\boldsymbol a}_j^k$, thereby saving computation cost.

It should also be noted that the algorithms presented in the following sections are solely based on the computed values of the primal and dual variables of ADMM. They do not need any information about the power system model parameters, nor the PMU measurements ${y}_i$. This is the main difference between our work and the work in \cite{tabuada}. In \cite{tabuada}, the authors derived detectability results based on the properties of the state matrix $\boldsymbol L$ in \eqref{Wang_model}. In our problem set-up, however, $\boldsymbol L$ consists of the Hankel matrices ${\hat {\boldsymbol H}}_i$ and ${\hat {\boldsymbol c}}_i$, both of which are filled with the measured outputs ${ y}_i$. The inherent assumption is that the central PDC does not have direct access to any ${ y}_i$, and therefore, does not know anything about the matrix $\boldsymbol L$. It only has access to the estimates ${\boldsymbol a}_i$ and the dual variables ${\boldsymbol w}_i$, $i=1,\dots,\,N$, and, therefore, must algorithmically figure out the detection and identification mechanisms based on these two variables only. This is the main contribution of the paper, compared to the model-based results of \cite{tabuada}.

\vspace{-0.1in}
\section{Data Manipulations with General Biases}\label{section3}

\subsection{Detecton of attacks}
For the ADMM algorithm stated in Section \ref{section2}, local PDCs only need to transmit the primal variable ${\boldsymbol a}_j^k$ to the central PDC. To make the algorithm attack-resilient, we next show that the central PDC can mandate all the local PDCs to transmit both ${\boldsymbol a}_j^k$ and ${\boldsymbol w}_j^{k-1}$.

We denote the value of ${\boldsymbol a}_i^k$ received at the central PDC at iteration $k$ as ${\bar{\boldsymbol a}}_i^k$. We define ${\bar{\boldsymbol a}}^k = [({\bar{\boldsymbol a}}^k_1)^{\rm T}, ({\bar{\boldsymbol a}}^k_2)^{\rm T}, \ldots, ({\bar{\boldsymbol a}}^k_N)^{\rm T}]^{\rm T}$. If $i \in \mathcal{S}$, ${\bar{\boldsymbol a}}_i^k = {\boldsymbol a}_i^k + \Delta_i^k$; otherwise, ${\bar{\boldsymbol a}}_i^k = {\boldsymbol a}_i^k$. 
If there is no attacker in the system, then at the first iteration $\frac{1}{N}\sum\limits_{i=1}^N {\bar{\boldsymbol a}}_i^1 = \frac{1}{N}\sum\limits_{i=1}^N{\boldsymbol a}_i^1 = {\boldsymbol z}^1$. According to \eqref{p3_1} then we have $\frac{1}{N} \sum\limits_i^N {\boldsymbol w}_i^j = {\boldsymbol 0}_{2n \times 1}$, $j = 0~or~1$, where ${\boldsymbol 0}_{2n \times 1}$ is a $2n \times 1$ matrix whose elements are all zeros; while if any PDC is biased, then ${\boldsymbol z}^1 = \frac{1}{N}\sum\limits_{i=1}^N{\bar {\boldsymbol a}}_i^1 = \frac{1}{N}\sum\limits_{i=1}^N{\boldsymbol a}_i^1 + \Delta^1$ and, hence, ${\frac{1}{N} \sum\limits_{i=1}^N {\boldsymbol w}_{i}^1} = -\rho \Delta^1$.
Thus the central PDC can detect the presence of malicious users at the second iteration by simply checking the average value of the dual variables. Next we will describe two algorithms by which central PDC can identify which local PDCs are malicious.

\subsection{S-ADMM for Identifying Malicious PDCs}\label{sec3}

For any pair of PDCs $(i, j)$, we define the quantity ${\boldsymbol d}_{i,j}^k={\bar{\boldsymbol a}}_i^k - {\boldsymbol a}_j^k$. For $i \in {\mathcal{S}}$ and $j \notin \mathcal{S}$, from \eqref{Wang_model} one can write
\begin{eqnarray}\label{diff_a}
{\boldsymbol{d}}_{i,j}^{k+1} &= & {\bar{\boldsymbol a}}_i^{k+1} - {\bar{\boldsymbol a}}_j^{k+1} \nonumber\\
&=& ({\boldsymbol L}_i - {\boldsymbol L}_j){\tilde{\boldsymbol a}}^k + ({\boldsymbol P}_i - {\boldsymbol P}_j)\Delta^k + \Delta_i^{k+1},
\end{eqnarray}
where $ {\tilde {\boldsymbol a}}^{k} = \left[ {\begin{array}{*{20}{c}}
{{{\boldsymbol a}^{k }}}\\
{{{\boldsymbol a}^{k-1 }}}
\end{array}} \right]$, and ${\boldsymbol L}_i$ and ${\boldsymbol P}_i$ are the $i^{th}$ $(2n \times 2n N)$ block rows of $\boldsymbol L$ and $\boldsymbol P$, respectively. On the other hand, for $i, j \notin \mathcal{S}$,
\begin{align}\label{diff_a2}
&{\boldsymbol{d}}_{i,j}^{k+1}  =  ({\boldsymbol L}_i - {\boldsymbol L}_j){\tilde{\boldsymbol a}}^k + ({\boldsymbol P}_i - {\boldsymbol P}_j)\Delta^k.
\end{align}
Comparing \eqref{diff_a} and \eqref{diff_a2}, it can be seen that if the minimum absolute value of a non-zero element of $\Delta_i^{k+1}$ is large enough, then the difference of estimates between two non-malicious PDCs can be much smaller compared to the difference between any malicious PDC and any non-malicious PDC. Thus, at any iteration the central PDC will be able to separate the incoming messages into at least two groups based of the values of the biases by simply computing the difference between every pair of messages arriving from the local PDCs. The messages without biases will belong to the same group. 
We define a threshold $\gamma_a^k$ to identify the group members at iteration $k$ as scalar threshold: 
\begin{align}\label{gamma}
\gamma_a^k = \min\bigg\{\frac{||{\bar{\boldsymbol a}}_{max}^k||  -||{\bar{\boldsymbol a}}_{min}^k||}{N}, \quad N(||{\bar{\boldsymbol a}}_{min2}^k|| -||{\bar{\boldsymbol a}}_{min}^k||) \bigg\},
\end{align}
where $||{\bar{\boldsymbol a}}_{max}^k||$, $||{\bar{\boldsymbol a}}_{min}^k|| $, and $||{\bar{\boldsymbol a}}_{min2}^k|| $ are the maximum, minimum, second minimum values of $||{\bar{\boldsymbol a }}_j^k||$, $1 \le j \le N$, respectively. The norm symbol here represents the Euclidian norm. In what follows, we will simply use the symbol $||\cdot||$ to represent Euclidian norm. If $|||{\bar{\boldsymbol a}}^k_j|| - ||{\bar{\boldsymbol a}}^k_i||| \le \gamma_a^k$, then the central PDC classifies the vectors ${\boldsymbol a}^k_j$ and ${\boldsymbol a}^k_i$ to be in the same group; otherwise, ${\boldsymbol a}^k_j$ and ${\boldsymbol a}^k_i$ are treated to be in the different groups.

Note that, $||{\bar{\boldsymbol a}}_i^{k}|| - ||{\bar{\boldsymbol a}}_j^{k}|| = ||{\boldsymbol a}_i^{k} + \Delta_i^{k}|| - ||{\boldsymbol a}_j^{k}||$, $i \in \mathcal{S}$ and $j \notin \mathcal{S}$. After a few calculations it can be easily shown that for successful localization, i.e., for making $||{\bar{\boldsymbol a}}_i^{k}|| - ||{\bar{\boldsymbol a}}_j^{k}|| >\gamma_a^k$, the bias $\Delta_i^k$ must satisfy 
\begin{small}
\begin{align}
&||\Delta_i^k||_{\infty} - \frac{\Delta_{max}^k}{N} > \frac{||{\boldsymbol a}_{max}^k|| - ||{\boldsymbol a}_{min}^k||}{N} + ||{\boldsymbol a}_j^k|| - ||{\boldsymbol a}_i^k||_{\infty}\label{require1}\\
&||\Delta_i^k||_{\infty} > N\big(||{\boldsymbol a}_{min2}^k|| - ||{\boldsymbol a}_{min}^k||\big) + ||{\boldsymbol a}_j^k|| - ||{\boldsymbol a}_i^k||_{\infty},\label{require2}
\end{align}
\end{small}
for all $k$, where $||\cdot||_{\infty}$ represents maximum norm.
Otherwise, if $i, j \notin \mathcal{S}$ $||{\bar{\boldsymbol a}}_i^{k}|| - ||{\bar{\boldsymbol a}}_j^{k}|| = ||{\boldsymbol a}_i^{k} || - ||{\boldsymbol a}_j^{k}||$, to guarantee $||{\bar{\boldsymbol a}}_i^{k}|| - ||{\bar{\boldsymbol a}}_j^{k}|| < \gamma_a^k$, $\Delta_{max}^k$ must satisfy
\begin{align}\label{require3}
||\Delta_{max}^k||_{\infty} > N \big( ||{\boldsymbol a}_i^k|| - ||{\boldsymbol a}_j^k||\big) + ||{\boldsymbol a}||_{min}^k - ||{\boldsymbol a}_{max}^k||_{\infty}.
\end{align}
If the biases satisfy the requirements as in \eqref{require1}-\eqref{require3}, then
S-ADMM can successfully identify the malicious PDCs by simply tracking the differences $|||{\bar{\boldsymbol a}}_i^k||-||{\bar{\boldsymbol a}}_j^k|||$. In reality, however, these lower bounds may not mean much since the fundamental rationale behind the detection and localization are all based on the quality of the estimates, which depend on the numerical magnitude of the measurements that are specific to that particular distance event.
Algorithm \ref{detecting_S-ADMM} summarizes the implementation of this simple method.

\begin{algorithm}
\caption{Identifying malicious PDCs injected with general biases using S-ADMM}


{\bf{Detection:}}

1) At $k=1$ every local PDC computes ${\boldsymbol a}_{j}^{k+1}$ in \eqref{p3_2} and ${\boldsymbol w}_{j}^{k}$ in \eqref{p3_1}, $j = 1, \ldots, N$, and transmits these two to the central PDC.

2) At $k=2$ the central PDC calculates $\frac{1}{N}\sum\limits_{i=1}^N {\boldsymbol w}_{ i}^1$. If $\frac{1}{N}\sum\limits_{i=1}^N {\boldsymbol w}_i^1 \ne {\boldsymbol 0}_{2n \times 1}$, it suspects that there exists one or more malicious PDCs in the system.

{\bf{Identification:}}

3) If Step 2 is positive, for all $k > 2$ the central PDC computes the difference $|||{\bar{\boldsymbol a}}_i^k|| -||{\bar{\boldsymbol a}}_j^k|| |$, $1 \le i, j \le N$, and the threshold $\gamma_a^k$. It then compares these differences to the threshold, and separates ${\bar {\boldsymbol a}}^k$ into groups.

4) The central PDC finds the index $j$ of the vectors ${\bar{\boldsymbol a}}_j^k$, $1 \le j \le N$, whose 2-norm is minimum. It then picks the group where the vector with this index is located, and classifies this group as unbiased.

5) The central PDC repeats this classification for a sufficiently large iteration $s$. If the identified non-malicious PDCs are consistent through these iterations, it finally confirms that these PDCs are unbiased.

6) Onwards from iteration $s+3$, the central PDC ignores any message coming from the malicious PDCs, and simply carries out S-ADMM with the remaining non-malicious PDCs using \eqref{p3}. The final solution of this S-ADMM will lead to the solution of \eqref{p1} as the LS problem is convex with $s+3$ being an initial iteration for the rest of the non-malicious S-ADMM.

\label{detecting_S-ADMM}
\end{algorithm}

{Since the success of this algorithm is contingent on the quality of the estimates, it is difficult to specify a universial lower bound for $s$ in Step 4. Generally, one would choose $s$ to be large, but it actual value must be decided on a case-by-case basis. Also, note that after identifying the attacked PDCs, Step 6 of the algorithm eliminates these PDCs, and continues normal estimation using the good PDCs. The reason why omitting a certain set of PDCs translates into retainment of stability is because the original least-squares problem in (5) is based on consensus. This means that if every node estimates ${\boldsymbol a}_i$ correctly, then the ADMM algorithm is guaranteed to converge to the centralized least-squares solution $\boldsymbol a$ \cite{boyd}. More importantly, the solution for this convergence does not depend on how many PDCs are there in the system. As every PDC is trying to reach the same optimal point $\boldsymbol a$, it does not matter whether there are $N$ PDCs, or less than $N$ PDCs. The speed of convergence, of course, may slow down as more and more PDCs are omitted, but the final solution will remain the same.}

\begin{figure}
\centering
\includegraphics[width=8.5cm,height=6.5cm]{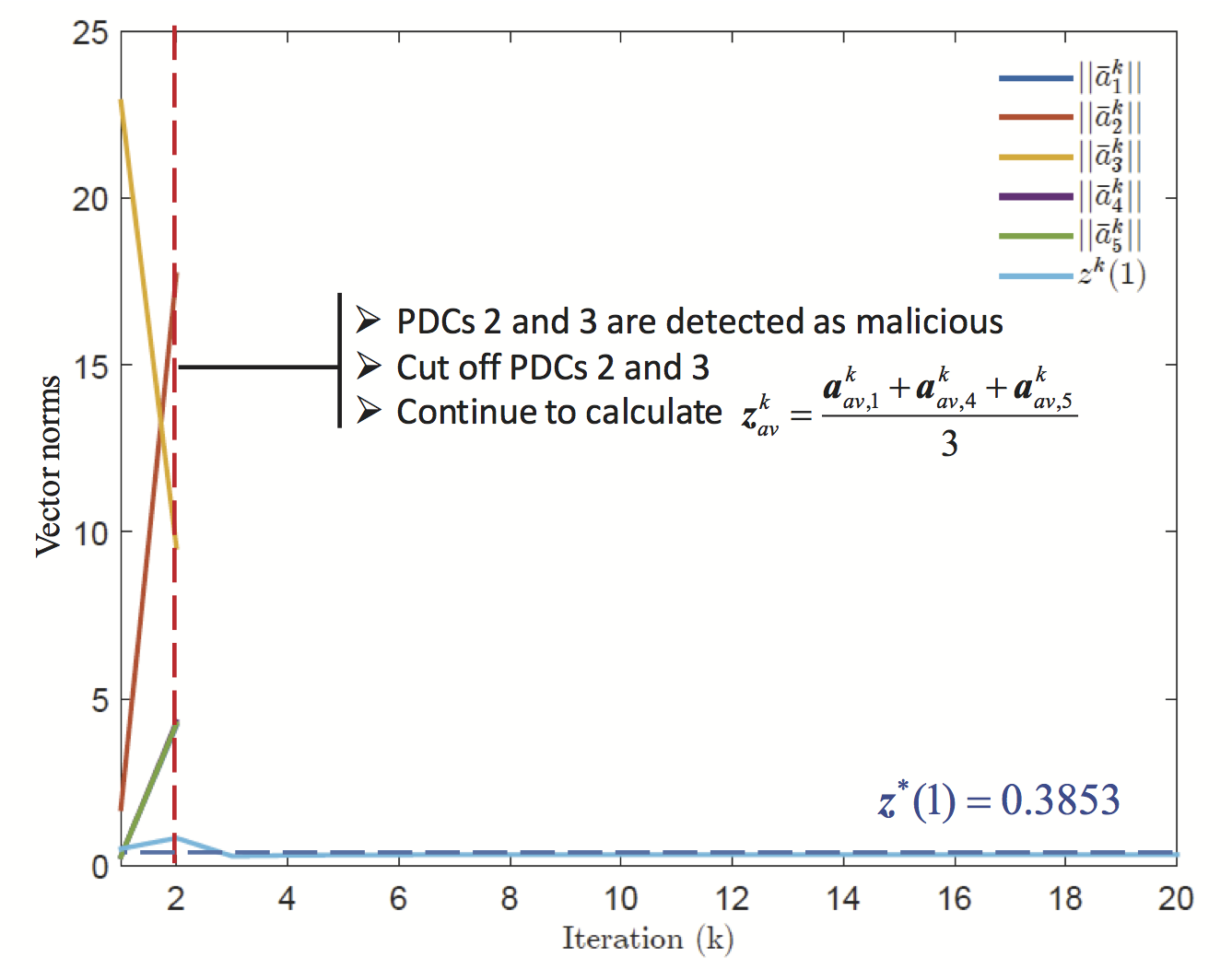}
\caption{ Evolution of the norms of the local estimates, and the average ${\boldsymbol z}^k$ before and after detection using S-ADMM. For convenience, we only show the first element of ${\boldsymbol z}^k$.}
\label{s_admm_dec_est_aft}
\end{figure}

We illustrate the approach with a simple example. Consider a system with 5 local PDCs. The second and third PDCs are respectively injected with $\Delta_2^k  = \delta_2^k {\boldsymbol 1}_{2n \times 1}$ and $\Delta_3^k  = \delta_3^k {\boldsymbol 1}_{2n \times 1}$, where $\delta_2^k $ and $\delta_3^k$ are two different arbitrary time-varying numbers and ${\boldsymbol 1}_{2n \times 1}$ is a $2n \times 1$ vector with all elements one. All other PDCs are unbiased. Fig. \ref{s_admm_dec_est_aft} shows the first element of the consensus vector ${\boldsymbol z}^k$ before and after detecting the malicious PDCs uing Algorithm 1. In the figure, it can be seen that at iteration $k=2$ we have $||{\bar{\boldsymbol a}}_1^2|| = 4.1864$, $||{\bar{\boldsymbol a}}_2^2|| = 17.7189$, $||{\bar{\boldsymbol a}}_3^2|| = 9.5428$, $||{\bar{\boldsymbol a}}_4^2|| = 4.3161$, $||{\bar{\boldsymbol a}}_5^2|| = 4.2459$. Thus, $||{\bar{\boldsymbol a}}^2_{max}|| = 17.7189$, $||{\bar{\boldsymbol a}}^2_{min}|| = 4.1864$, and $||{\bar{\boldsymbol a}}^2_{min2}|| = 4.2459$. The threshold, therefore, is $\gamma_a^2 =0.2975$. The estimates ${\bar{\boldsymbol a}}^2$ are separated into three groups: ${\bar{\boldsymbol a}}_1^2$, ${\bar{\boldsymbol a}}_4^2$, and ${\bar{\boldsymbol a}}_5^2$ are in the first group, ${\bar{\boldsymbol a}}_2^2$ and ${\bar{\boldsymbol a}}_3^2$ are in the second and the third groups, respectively. The magnitude $||{\bar{\boldsymbol a}}_1^2||$ in the first group is minimum. Thus, PDCs 1, 4 and 5 are identified as non-malicious, or alternatively, PDCs 2 and 3 are identified as malicious. After iteration $2$, the central PDC cuts off communication with PDCs 2 and 3, and only calculates the average of messages received from the unbiased PDCs, leading to ${\boldsymbol z}^k = \frac{1}{3}({\boldsymbol a}_{1}^k +{\boldsymbol a}_{4}^k + {\boldsymbol a}_{5}^k)$, $\forall k>2$. The estimates thereby asymptotically converge to the ideal solution ${\boldsymbol z}^*$, as expected.

\subsection{Round-Robbin ADMM for Detecting the Malicious Users}

\begin{table*}[h!]
\centering
\caption{Comparison of S-ADMM and RR-ADMM for detection of attacks with large biases}
\begin{tabular}{c|c| c |c|c}
\hline
& Variable used for detection & Minimum \# of computations per iteration & Minimum \# of iterations needed to identify & Bias Magnitude \\
\hline
S-ADMM & ${\boldsymbol a}_{av, j}^k$ & $\frac{N(N-1)}{2}$ & 1 & Less stringent \\
\hline
RR-ADMM & ${\boldsymbol z}_{rr}^k$ & $1$ & $k_{min}+N$ & More stringent \\
\hline
\end{tabular}
\label{suma}
\end{table*}

Algorithm 1, though very simple to implement, may suffer from computational challenges as the central PDC has to process $2nN$ number of elements at every iteration $k$. In practice, there may be thousands of virtual computers installed in a cloud network serving the purpose of these local PDCs, as shown in our recent paper \cite{yufeng}, which makes $N$ a very large number. The computational memory of the central PDC may not be able to track such a large number of variables per iteration in real-time. Instead it may only want to track or {\it keep an eye} on a much smaller number of variables such as only the consensus vector ${\boldsymbol  z}^k$ which has $2n$ number of elements in it. Under that condition, if any local estimate ${\boldsymbol  a}_i^k$ is corrupted by bias $\Delta_i^k$ at every iteration, then it will be impossible for the central PDC to identify the malicious PDCs, or identify which PDCs are unbiased, just by tracking ${\boldsymbol  z}^k$. The main question, therefore, is how can the central PDC catch the manipulators by simply tracking ${\boldsymbol  z}^k$ over every iteration? We next propose a variant of (\ref{p3}) using a Round-Robin strategy replacing the averaging step (\ref{p3_3}) to solve this problem. We refer to this algorithm as Round-Robin ADMM or RR-ADMM in short.

The basic strategy for RR-ADMM is as follows. At iteration $k=1$, for example, the central PDC receives local estimates ${\boldsymbol a}_j^1$ from every local PDC, but computes ${\boldsymbol z}^1$ simply as ${\boldsymbol z}^1= \alpha {\boldsymbol a}_1^1$, where $\alpha$ is a constant non-zero number. Then the central PDC sends ${\boldsymbol z}^1$ back to the local PDCs following Step 4 of S-ADMM.
Similarly, in iteration $k=2$, the central PDC uses ${\boldsymbol z}^2=\alpha {\boldsymbol a}_2^2$, at iteration $k=3$ it uses ${\boldsymbol z}^3= \alpha {\boldsymbol a}_3^3$, and so on. In general, ${\boldsymbol z}^k=\alpha {\boldsymbol a}^k_{((k-1)\bmod N) + 1}$. $N$ successive iterations constitute one {\it period} of RR-ADMM where $N$ is the total number of local PDCs. After $N$ iterations, the central PDC will again start from PDC 1, then PDC 2, and so on. For convenience of expression, we denote the consensus variables of RR-ADMM and S-ADMM at iteration $k$ as ${\boldsymbol z}_{rr}^k$ and ${\boldsymbol z}_{av}^k$, the latter being updated in \eqref{p3_3}. Similarly, the local PDC estimates and its dual variable will be denoted as ${\boldsymbol a}_{rr, j}^k$ and ${\boldsymbol w}_{rr, j}^k$, and ${\boldsymbol a}_{av,j}^k$ as in \eqref{p3_2} and ${\boldsymbol w}_{av, j}^k$ as in \eqref{p3_1}, $1 \le j \le N$, respectively.

{\bf Remark}: Note that the purpose of using RR-ADMM is only to detect the malicious local PDC, not for obtaining the optimal solution of \eqref{p1}. This is because this algorithm is run by the central PDC stealthily, while every local PDC still believes that the central PDC uses S-ADMM to calculate ${\boldsymbol z}^k$, and thereby updates ${\boldsymbol a}_{rr,j}^{k+1}$ in the same way as in \eqref{p3_2}. Therefore, this algorithm should be treated more as a S-ADMM with a RR-averaging step, rather than a true RR-ADMM where every step of (\ref{p3}) would have to be modified in accordance to the RR strategy.

Considering some of the PDCs to be malicious, the ADMM update equations using RR-averaging can be written as:
\begin{subequations}
\label{r3}
\begin{align}
{\boldsymbol{w}}_{rr, i}^{k}=&{\boldsymbol{w}}_{rr, i}^{k-1}+\rho({\boldsymbol{a}}_{rr,i}^{k}-{\boldsymbol{z}}^{k}_{rr}), \label{r3_1}\\
{\boldsymbol{a}}_{rr,i}^{k+1}=&((\hat{\boldsymbol{H}}^{k}_i)^T\hat{\boldsymbol{H}}^{k}_i+\rho {\boldsymbol{I}})^{-1}((\hat{\boldsymbol{H}}^{k}_i)^T\hat{\boldsymbol{c}}^{k}_i-{\boldsymbol{w}}^{k}_{i}+\rho {\boldsymbol{z}}_{rr}^{k}), \label{r3_2}\\
{\boldsymbol{z}}_{rr}^{k+1}=&\alpha {\bar{\boldsymbol a}}^{k+1}_{rr, ((k-1)\bmod N) + 1}\nonumber\\
=& \alpha ({\boldsymbol a}^{k+1}_{rr,((k-1)\bmod N) + 1} + \Delta^{k+1}_{((k-1)\bmod N) + 1}) . \label{r3_3}
\end{align}
\end{subequations}
From \eqref{r3_3}, it is clear that unlike S-ADMM where ${\boldsymbol z}_{av}^k$ only depends on the average value of the biases, the expression for ${\boldsymbol z}_{rr}^k$ is influenced by the injection of the ${[((k-1)\bmod N) + 1]}^{th}$ PDC, i.e., $\Delta^k_ {((k-1)\bmod N) + 1}$, which provides a potential `signature' to catch the identities of the PDCs with disturbances. In \cite{LiaoA16}, we proposed a way to detect the malicious PDC by finding the maximum $\xi^i = |||{\boldsymbol z}_{rr}^i|| - ||{\boldsymbol z}_{rr}^{i+1}|||+|||{\boldsymbol z}_{rr}^i||- ||{\boldsymbol z}_{rr}^{i-1}|||$, $1 <i < N+1$, which holds only when there is exactly one malicious PDC. We next provide an alternative algorithm that applies for multiple attacks.

From \eqref{r3}, at iteration $k$, we can write the consensus vector ${\boldsymbol z}_{rr}^k$ as
\begin{align}\label{eq:zrriter}
&{\boldsymbol z}_{rr}^k = \alpha \Delta_b^k + \alpha \left(({\hat {\boldsymbol H}}_b^{k-1})^{\rm T}({\hat {\boldsymbol H}}_b^{k-1}) + \rho {\boldsymbol I}_{2n}\right)^{-1} \times \nonumber\\
&\left[({\hat {\boldsymbol H}}_b^{k-1})^{\rm T}{\hat {\boldsymbol c}}_b^{k-1} - {\boldsymbol w}_{rr, b}^0 - \rho \sum\limits_{j = 1}^{k-1}({\boldsymbol a}_{rr,b}^j - {\boldsymbol z}_{rr}^j)+\rho{\boldsymbol z}_{rr}^{k-1}\right],
\end{align}
where $b = \left({(k-1)}\bmod {N}\right) + 1$. We assume that the minimum absolute value of the element in the variable $\Delta_b^k$ is large enough. It follows from \eqref{eq:zrriter} that the minimum value of $||{\boldsymbol z}_{rr}||$ in one period must be from a non-malicious PDC. Let $k_{min}$ denote the iteration index in the period $k=1,\dots, N$ where the magnitude of $||{\boldsymbol z}_{rr}^k||$ is minimum, $1 \le k \le N$, and define a threshold $\gamma_z$ as
\begin{align}
\gamma_z =||{\boldsymbol z}_{rr}^{k_{min}+N}|| - ||{\boldsymbol z}_{rr}^{k_{min}}||.
\end{align}
If $||{\boldsymbol z}_{rr}^{k}||>||{\boldsymbol z}_{rr}^{k_{min}}|| + \gamma_z$, $k_{min} \le k \le k_{min}+N-1$, then the central PDC infers that the $[(k-1)\bmod N + 1]^{th}$ PDC is attacked. Algorithm \ref{detecting} summarizes this detection mechanism.

\begin{algorithm}\label{detecting}
\caption{Identifying malicious PDC with general biases using RR-ADMM}


{\bf{Detection:}}

{1) At $k = 1$ every local PDC computes ${\boldsymbol a}_{j}^{k+1}$ in \eqref{p3_2} and ${\boldsymbol w}_{j}^{k}$ in \eqref{p3_1}, $j = 1, \ldots, N$, and transmits these two to the central PDC.}

{2) At $k = 2$ the central PDC calculates $\frac{1}{N}\sum\limits_{i=1}^N {\boldsymbol w}_{av, i}^1$. If $\frac{1}{N}\sum\limits_{i=1}^N {\boldsymbol w}_i^1 \ne {\boldsymbol 0}_{2n \times 1}$, it suspects that there exists one or more malicious PDCs in the system.}

{\bf{Identification:}}

{3) If Step 2 is positive, for $k>2$ the central PDC switches to RR-ADMM. That is, every local PDC computes ${\boldsymbol a}_{rr, j}^{k+1}$ in \eqref{r3_2} and ${\boldsymbol w}_{rr, j}^{k+1}$ in \eqref{r3_1}, $j = 1, \ldots, N$, and transmits them to the central PDC.}

4) When $k \ge N$, the central PDC searches for the minimum value $||{\boldsymbol z}_{rr}^{k_{min}}||$ in one period and its iteration index $k_{min}$.

5) Waiting till $k \ge k_{min}+N$, the central PDC computes the threshold $\gamma_z =||{\boldsymbol z}_{rr}^{k_{min}+N}|| - ||{\boldsymbol z}_{rr}^{k_{min}}||$. It then compares $||{\boldsymbol z}_{rr}^{i}||$ to $||{\boldsymbol z}_{rr}^{k_{min}}|| + \gamma_z$ for $k_{min}\le i \le k_{min}+N-1$.

6) If $||{\boldsymbol z}_{rr}^{i}||>||{\boldsymbol z}_{rr}^{k_{min}}|| + \gamma_z$, then the central PDC identifies the ${[(i - 1)\bmod N + 1]}^{th}$ PDC to be malicious.

7) The central PDC repeats this classification for a few iteration, say up to iteration $s$. If the identified non-malicious PDCs are consistent through these iterations, it finally confirms that these PDCs are unbiased.

8) Onwards from iteration $s+3$, the central PDC ignores any message coming from the malicious PDCs, and simply carries out S-ADMM with the remaining non-malicious PDCs using \eqref{p3}. The final solution of this S-ADMM will lead to the solution of \eqref{p1} as the LS problem is convex with $s+3$ being an initial iteration for the rest of the non-malicious S-ADMM.

\label{detecting}
\end{algorithm}
Notice that if the biases are constant, the central PDC does not need to search for the minimum value of $||{\boldsymbol z}_{rr}^k||$, $1 \le k \le N$, and the threshold $\gamma_z$ can be defined as $||{\boldsymbol z}_{rr}^{N+1}|| -||{\boldsymbol z}_{rr}^{1}||$. In that case, the central PDC will only wait for $N+1$ iterations, not $k_{min}+N$ iterations.

\begin{figure}
\centering
\includegraphics[width=8.5cm,height=6.5cm]{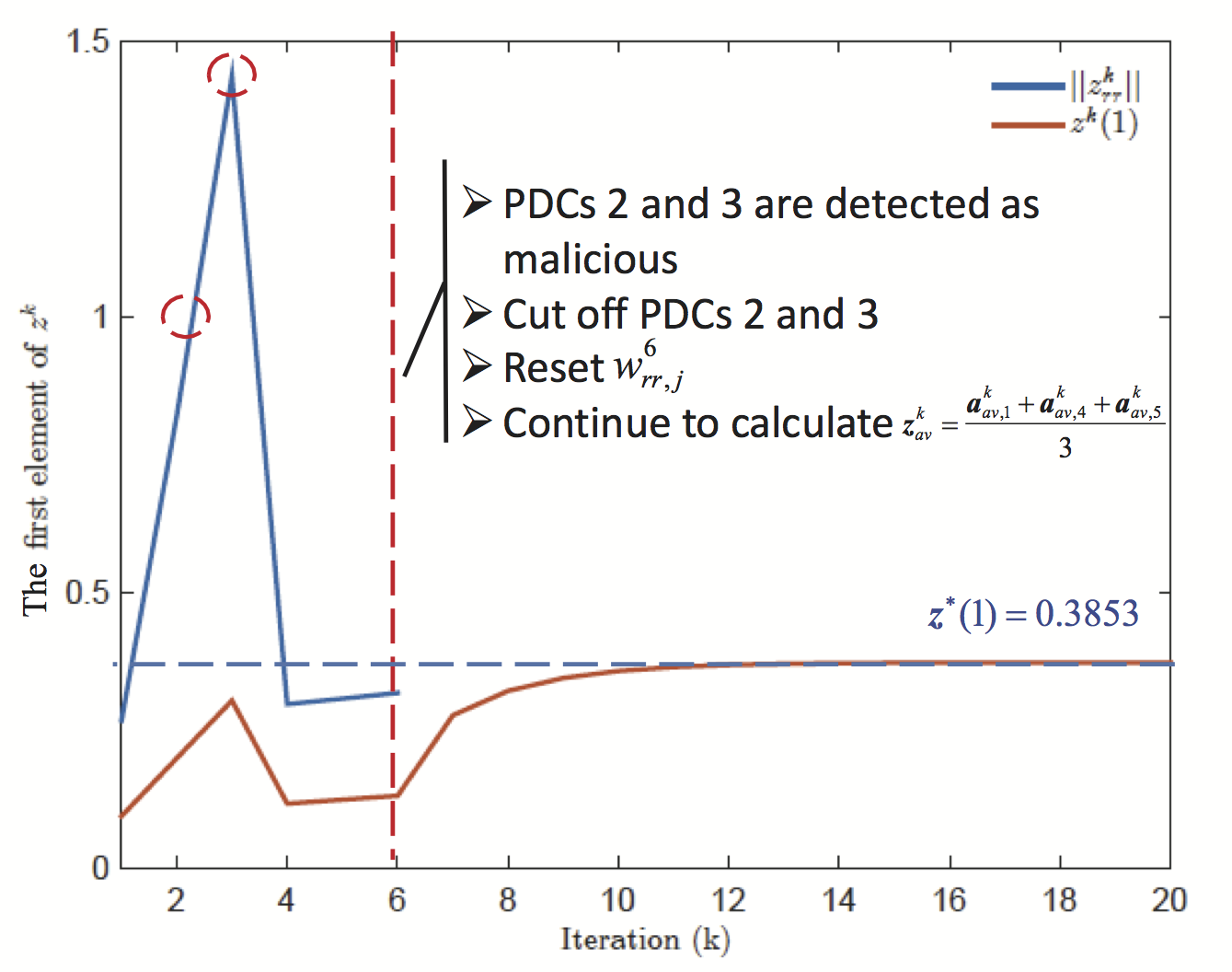}
\caption{Evolution of the average ${\boldsymbol z}^k$ before and after detection using RR-ADMM. For convenience, we only show the first element of ${\boldsymbol z}^k$.}
\label{rr_w1_det_s_dec_aft}
\end{figure}

Consider the same example as in Fig. \ref{s_admm_dec_est_aft}. Fig. \ref{rr_w1_det_s_dec_aft} shows the first element of the consensus vector ${\boldsymbol z}_{rr}^k$ before and after detection using RR-ADMM. In the first period, $||{\boldsymbol z}_{rr}^1|| = 0.2672$, $||{\boldsymbol z}_{rr}^2|| = 0.8192$, $||{\boldsymbol z}_{rr}^3|| = 1.4356$, $||{\boldsymbol z}_{rr}^4|| = 0.2964$, and $||{\boldsymbol z}_{rr}^5|| = 0.3064$. Thus, $||{\boldsymbol z}_{rr}^1||$ is minimum and $k_{min} = 1$. The threshold is computed as $\gamma_z = ||{\boldsymbol z}_{rr}^6|| - ||{\boldsymbol z}_{rr}^1|| = 0.0497$. Only $||{\boldsymbol z}_{rr}^2||$ and $||{\boldsymbol z}_{rr}^3||$ are larger than $\gamma_z + ||{\boldsymbol z}_{rr}^1||$. So PDCs 2 and 3 are identified as malicious. After detecting these manipulators at iteration 6, the central PDC cuts off the signal from  PDCs 2 and 3, and only calculates the average of messages received from the other local PDCs leading to ${\boldsymbol z}_{av}^k = \frac{1}{3}({\boldsymbol a}_{av, 1}^k +{\boldsymbol a}_{av, 4}^k + {\boldsymbol a}_{av, 5}^k)$, $\forall k>6$. Note that at the same time the dual variable ${\boldsymbol w}_j^k$ must be reset to its initial value. At iteration 7, ${\boldsymbol a}_{av,j}^7$ is updated using the value of ${\boldsymbol z}_{rr}^6$.

The requirement on the magnitude of the biases for successful detection (i.e. for minimizing false positives) is much more strict for RR-ADMM than for S-ADMM. This is because the threshold $\gamma_z$ is dependent on the speed of divergence of the elements of ${\boldsymbol z}_{rr}$ in one period. For S-ADMM, however, ${\bar{\boldsymbol a}}_{av, j}^k$, $1 \le j \le N$ are compared only at one iteration, and so the biases are less affected by the speed of divergence.
Also note that if every element of the bias vector at every iteration is non-zero, and the central PDC knows this information, then the detection can be done by only using any chosen element of ${\bar {\boldsymbol a}}_{av, j}^k$, $1 \le j \le 2n$ and ${\boldsymbol z}_{rr}^k$ (for S- and RR-ADMM respectively) instead of using the vector norms.
Table \ref{suma} compares S-ADMM and RR-ADMM for detecting attacks with large magnitudes.


\subsection{Random Order of RR-ADMM}

Note that the order of choosing the PDCs for the averaging step in the RR-ADMM need not be fixed. The algorithm should succeed with any random order as long as every PDC is visited exactly once in a period. For example, if there are 4 local PDCs, then the central PDC need not follow the order as (1, 2, 3, 4) for every period. It can, for instance, choose (1, 2, 3, 4) for the first period, (1, 3, 4, 2) for the second, (2, 4, 3, 1) for the third, and so on. This gives the central PDC the flexibility in choosing ${\boldsymbol z}^k_{rr}$ in case any of the local estimates do not arrive on time due to message loss or denial-of-service. In that case, Algorithm \ref{detecting} needs to be modified slightly to accommodate this random order. After Step 3 in Algorithm \ref{detecting}, the central PDC should find the index of the local PDC corresponding to the iteration index $k_{min}$. Let this PDC index be $m$. Let $k^\ast$ be the iteration index between $k=N+1$ and $k=2N$, considering ${\boldsymbol z}^{k^\ast}_{rr}=\alpha{\bar{\boldsymbol a}}^{k^\ast}_ {rr,m}$. The threshold is then changed to $\gamma_{z} = ||{\boldsymbol z}_{rr}^{k*}||-||{\boldsymbol z}_{rr}^{k_{min}}||$. After that the central PDC compares $||{\boldsymbol z}_{rr}^k||$ to $\gamma_{z} + ||{\boldsymbol z}_{rr}^{k_{min}}||$, $1 <k < N$. If $||{\boldsymbol z}_{rr}^k||> \gamma_{z} + ||{\boldsymbol z}_{rr}^{k_{min}}||$ and if ${\boldsymbol z}_{rr}^k = \alpha {\bar{\boldsymbol a}}_{rr, i^*}^k$, then the ${i^*}^{th}$ PDC is identified as malicious.

We will illustrate this approach using a real power system simulation in Case 5 of Section \ref{simulation}.

\section{Data Manipulations with Small Biases}

\subsection{S-ADMM for Detecting Malicious Users with Small Biases}

The basic approach for this method is the same as in Section \ref{sec3}. Recall equations (\ref{diff_a}) and (\ref{diff_a2}) as follows. {For $i \in \mathcal{S}$ and $j \notin \mathcal{S}$}
\begin{eqnarray}\label{diff_a3}
{\boldsymbol{d}}_{i,j}^{k+1} =({\boldsymbol L}_i - {\boldsymbol L}_j){\tilde{\boldsymbol a}}^k + ({\boldsymbol P}_i - {\boldsymbol P}_j)\Delta^k + \Delta_i^{k+1},
\end{eqnarray}
while {if $i, j \notin \mathcal{S}$, then}
\begin{align}\label{diff_a4}
{\boldsymbol{d}}_{i,j}^{k+1} & =  ({\boldsymbol L}_i - {\boldsymbol L}_j){\tilde{\boldsymbol a}}^k + ({\boldsymbol P}_i - {\boldsymbol P}_j)\Delta^k \nonumber \\
 &= ({\boldsymbol L}_i - {\boldsymbol L}_j){\tilde{\boldsymbol a}}^k + \rho({A}_i - {A}_j)\Delta^k
\end{align}
where the last equation follows from the definition of $\boldsymbol P$ in Section \ref{sec2}. The problem, however, is that if $||\Delta_i^{k+1}||_{\infty}$ is small, then the value of $ ||\rho ({A}_j - {A}_i)\Delta^k + \Delta_i^{k+1}||$ in (\ref{diff_a3}) may become comparable to $||\rho ({A}_i - {A}_j)\Delta^k||$ in (\ref{diff_a4}), thereby leading to incorrect classification. One way to bypass this would be to reduce the value of $\rho >0$ such that the difference between the LHS of (\ref{diff_a3}) and (\ref{diff_a4}) is still large enough for detection despite $||\Delta_i^{k+1}||_{\infty}$ being a small number. Thus, the only difference of this approach from that in Section \ref{sec3} is that the ISO must ask every local PDC to reduce their penalty factor $\rho$ once it realizes the presence of a false-data injector. Algorithm \ref{detecting_S-ADMM_small} describes this method.

\begin{figure}
\centering
\includegraphics[width=9cm,height=7cm]{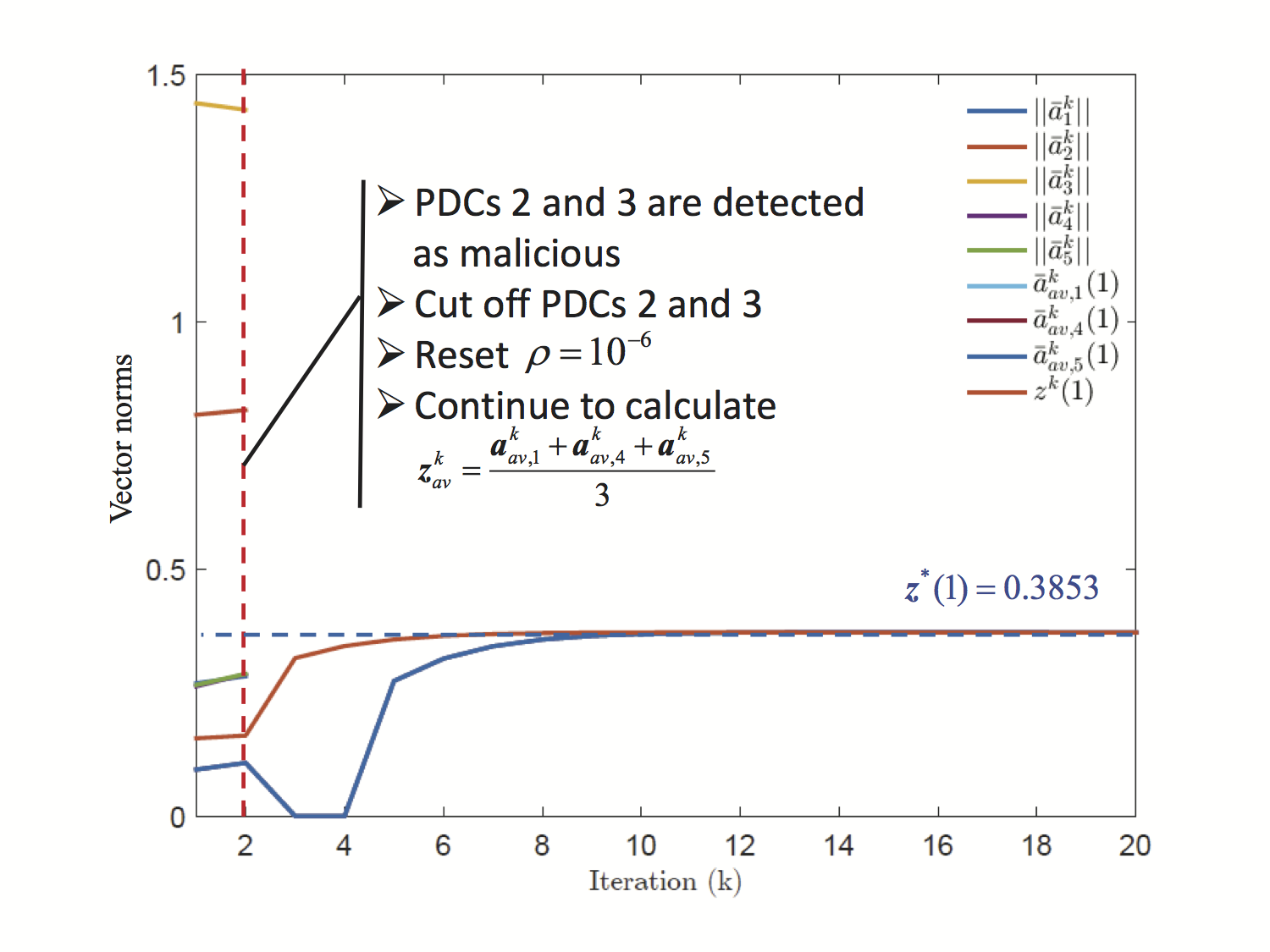}
\caption{Evolution of the norms of the local estimates, and the average ${\boldsymbol z}_{av}^k$ before and after detection of attack with small biases using S-ADMM. For convenience, we only show the first element of ${\boldsymbol z}_{av}^k$.}
\label{s_dec_est_after}
\end{figure}

\begin{algorithm}
\caption{Identifying malicious PDCs with small biases using S-ADMM}

{\bf{Detection:}}

{1) At $k=1$ every local PDC computes ${\boldsymbol a}_{j}^{k+1}$ in \eqref{p3_2} and ${\boldsymbol w}_{j}^{k}$ in \eqref{p3_1}, $j = 1, \ldots, N$, and transmits these two to the central PDC.}

{2) At $k=2$ the central PDC calculates $\frac{1}{N}\sum\limits_{i=1}^N {\boldsymbol w}_{av, i}^1$. If $\frac{1}{N}\sum\limits_{i=1}^N {\boldsymbol w}_i^1 \ne {\boldsymbol 0}_{2n \times 1}$, it suspects that there exists one or more malicious PDCs in the system.}

{\bf{Identification:}}

{3) If Step 2 is positive, for $k>2$ the central PDC asks every individual PDC $j$ to update ${\boldsymbol a}_{av, j}^{k+1}$ as in \eqref{p3_2} by reducing the value of $\rho$, $j = 1, \ldots, N$.}

4) Every local PDC updates $\rho$, and computes ${\boldsymbol a}_{av, j}^{k+1}$ in \eqref{p3_2}, ${\boldsymbol z}_{av}^{k+1}$ in \eqref{p3_3}, and ${\boldsymbol w}_{av, j}^{k+1}$ in \eqref{p3_1}, $j = 1, \ldots, N$, and transmits them to the central PDC.

5) For all $k > 2$, the central PDC computes the difference $|||{\bar{\boldsymbol a}}_{av, i}^k|| -||{\bar{\boldsymbol a}}_{av,j}^k|| |$, $1 \le i, j \le N$, and the threshold $\gamma_a^k$. It then compares these differences to the threshold, and separates ${\bar{\boldsymbol a}}_{av}^k$ into groups.

6) The central PDC finds the index $j$ of the vectors ${\bar{\boldsymbol a}}_{av, j}^k$, $1 \le j \le N$, whose 2-norm is minimum. It then picks the group where the vector with this index is located, and classifies the group as unbiased.

7) The central PDC repeats this classification for a few iteration, say up to iteration $s$. If the identified non-malicious PDCs are consistent through these iterations, it finally confirms that these PDCs are unbiased.

8) Onwards from iteration $s+3$, the central PDC ignores any message coming from the malicious PDCs, resets the value of $\rho$ to its initial value, and simply carries out S-ADMM with the remaining non-malicious PDCs using \eqref{p3}. The final solution of this S-ADMM will lead to the solution of \eqref{p1} as the LS problem is convex with $s+3$ being an initial iteration for the rest of the non-malicious S-ADMM.

\label{detecting_S-ADMM_small}
\end{algorithm}

{Notice that the penalty factor $\rho$ is a private parameter, and hence unknown to the attacker. However, following the attack model stated in Section \ref{sec2}, the communication link connecting the central PDC to any of the local PDCs, is assumed to be uncompromised. Thus, when the central PDC broadcasts the instruction to reduce $\rho$, then every PDC whether attacked or unattacked, must be able to follow this instruction. This satisfies Step 3 of Algorithm \ref{detecting_S-ADMM_small}.}

Consider the same example as in Section \ref{section3}. As before, we assume there are five local PDCs and one central PDC. The second and the third local PDCs are malicious, i.e., they send their estimates with two different arbitrary biases $\Delta_2^k$ and $\Delta_3^k$ at iteration $k$, respectively. Fig. \ref{s_dec_est_after} shows the trajectories of the first element of the vector ${\boldsymbol z}_{av}^k$ and the first elements of the un-biased vectors ${\bar {\boldsymbol a}}_{av, 1}^k$, ${\bar {\boldsymbol a}}_{av, 4}^k$, and ${\bar {\boldsymbol a}}_{av, 5}^k$ pre and post detection using S-ADMM. In Fig. \ref{s_dec_est_after}, at the second iteration, with $\rho$ reduced to $10^{-9}$, $||{\bar{\boldsymbol a}}_{av, 1}^2|| = 0.283$, $||{\bar{\boldsymbol a}}_{av, 2}^2|| = 0.8218$, $||{\bar{\boldsymbol a}}_{av, 3}^2|| = 1.4296$, $||{\bar{\boldsymbol a}}_{av, 4}^2|| = 0.2873$, $||{\bar{\boldsymbol a}}_{av, 5}^2|| = 0.2878$. The threshold is $\gamma_a^2 = 0.00215$. ${\bar{\boldsymbol a}}_{av}^2 $ is separated into three groups: ${\bar{\boldsymbol a}}_{av, 1}^2$, ${\bar{\boldsymbol a}}_{av, 4}^2$, and ${\bar{\boldsymbol a}}_{av, 5}^2$ are in the first group, ${\bar{\boldsymbol a}}_{av, 2}^2$ and ${\bar{\boldsymbol a}}_{av, 3}^2$ are in the second and the third groups, respectively. The value of $||{\bar{\boldsymbol a}}_{av, 4}^2||$ in the first group is minimum. Therefore, PDCs 1, 4, and 5 are identified to be non-malicious which matches with the true situation. After the second iteration, the central PDC cuts off communication with PDCs 2 and 3, and only calculates the average of messages received from the other local PDCs leading to ${\boldsymbol z}_{av}^k = \frac{1}{3}({\boldsymbol a}_{av, 1}^k +{\boldsymbol a}_{av, 4}^k + {\boldsymbol a}_{av, 5}^k)$, $k>2$. Notice that at iteration $3$, ${\rho}$ should be reset to $10^{-6}$. The estimates asymptotically converge to the true solution ${\boldsymbol z}^\ast$, as expected.

\subsection{RR-ADMM for Detecting Malicious Users with Small Biases}

The basic requirement for successful identification using S-ADMM is that the bias, no matter how small, should be large enough so that the difference of estimates between a malicious PDC and non-malicious PDC is notably larger than that between two non-malicious PDCs. In this section, we employ RR-ADMM using the dual variable ${\boldsymbol w}_{rr, j}^k$ instead of the consensus variable ${\boldsymbol z}_{rr}^k$ for detection where this requirement is less stringent.
From \eqref{r3_1} and \eqref{r3_3}, the difference ${\boldsymbol w}_{rr, i}^k - {\boldsymbol w}_{rr, i}^{k-1}$, considering $\alpha=1$, can be written as
\begin{align}\label{diff_w}
&{\boldsymbol w}_{rr, i}^k - {\boldsymbol w}_{rr, i}^{k-1}\nonumber\\
&= \rho({\boldsymbol a}_{rr, i}^k - {\boldsymbol a}_{rr, (k-1)\bmod N+1}^k - \Delta_{(k-1)\bmod N+1}^k).
\end{align}
If $1 \le k \le N$ and $i = k$, then \eqref{diff_w} is reduced to
\begin{align}\label{diff_ww}
{\boldsymbol w}_{rr, k}^k - {\boldsymbol w}_{rr, k}^{k-1} = \rho({\boldsymbol a}_{rr, k}^k - {\boldsymbol a}_{rr, k}^k - \Delta_{k}^k) = -\rho \Delta_k^k.
\end{align}
Therefore, if $\Delta_{k}^k ={\boldsymbol 0}_{2n \times 1}$, then ${\boldsymbol w}_{rr, k}^k - {\boldsymbol w}_{rr, k}^{k-1} = {\boldsymbol 0}_{2n \times 1}$, where ${\boldsymbol 0}_{2n \times 1}$ is a $2n \times 1$ vector with all zero elements. This indicates that if the $i^{th}$ PDC is not injected with any bias, then at the $i^{th}$ iteration, ${\boldsymbol w}_{rr, i}^i - {\boldsymbol w}_{rr, i}^{i-1} = {\boldsymbol 0}_{2n \times 1}$; if this condition is not met, the $i^{th}$ PDC must be malicious. The central PDC can check for the above inequality to identify the malicious PDC. For instance, at iteration 2, the central PDC should calculate the difference ${\boldsymbol w}_{rr, 2}^2 - {\boldsymbol w}_{rr, 2}^1$. If the difference is not equal to zero, PDC 2 must be malicious. At iteration 3, the central PDC should check the difference ${\boldsymbol w}_{rr, 3}^3 - {\boldsymbol w}_{rr, 3}^2 $; at iteration $N+1$, the central PDC should check the difference ${\boldsymbol w}_{rr, 1}^{N+1} - {\boldsymbol w}_{rr, 1}^N $. Thus, after $N+1$ iterations, the central PDC will be able to detect the identities of all manipulators no matter what the values of their biases are. Notice that for the exact accuracy of identification, the central PDC should compare each element of ${\boldsymbol w}_{rr, k}^k$ to the corresponding elements of ${\boldsymbol w}_{rr, k}^{k-1}$. If the central PDC compares the norms $||{\boldsymbol w}_{rr, k}^k||$ to $||{\boldsymbol w}_{rr, k}^{k-1}||$, then it is possible that under the condition $\Delta_{(k-1)\bmod N+1}^k \ne {\boldsymbol 0}_{2n \times 1}$, these two norms may turn out to be equal thereby making the identification inaccurate. The upshot is that every local PDC now must also send the dual variable to the central PDC at every iteration, thereby increasing the volume of the transmitted messages.
It must be noted, again, that the RR-ADMM should only be used for catching data-manipulators, and not to solve the optimal solution of \eqref{p1}.
Algorithm \ref{detection_w2} summarizes the RR-ADMM strategy.

\begin{algorithm}
\caption{Identifying malicious PDCs with small biases using RR-ADMM}


{\bf{Detection:}}

{1) At $k=1$ every local PDC computes ${\boldsymbol a}_{j}^{k+1}$ in \eqref{p3_2} and ${\boldsymbol w}_{j}^{k}$ in \eqref{p3_1}, $j = 1, \ldots, N$, and transmits these two to the central PDC.}

{2) At $k=2$ the central PDC calculates $\frac{1}{N}\sum\limits_{i=1}^N {\boldsymbol w}_{av, i}^1$. If $\frac{1}{N}\sum\limits_{i=1}^N {\boldsymbol w}_i^1 \ne {\boldsymbol 0}_{2n \times 1}$, it suspects that there exists one or more malicious PDCs in the system.}

{\bf{Identification:}}

{3) If Step 2 is positive, for $k>2$ the central PDC calculates the consensus variable ${\boldsymbol z}_{rr}^k$ using RR-ADMM instead of S-ADMM.}

4) Every local PDC updates ${\boldsymbol a}_{rr}^{k}$ in \eqref{r3_2},  ${\boldsymbol z}_{rr}^{k}$ in \eqref{r3_3}, and ${\boldsymbol w}_{rr}^k$ in \eqref{r3_1}, $j = 1, \ldots, N$, and transmits them to the central PDC.

5) After $N+1$ iterations, the central PDC calculates the differences ${\boldsymbol w}_{rr, i}^i - {\boldsymbol w}_{rr, i}^{i-1}$, $2 \le i \le N$, and the difference ${\boldsymbol w}_{rr, 1}^{N+1} - {\boldsymbol w}_{rr, 1}^{N}$. It then checks if these differences are exactly zero or not. PDCs showing non-zero differences are classified as malicious.

6) The central PDC repeats this classification for a few iteration, say up to iteration $s$. If the identified malicious PDCs are consistent through these iterations, it finally confirms these PDCs are malicious.

7) Onwards from iteration $ s+3$, the central PDC ignores any message coming from the malicious PDC, resets the dual variable $\boldsymbol w^{s + 3}_j$ to its initial value, and simply carries out S-ADMM with the remaining non-malicious PDCs using \eqref{p3}. The final solution of this S-ADMM will lead to the solution of \eqref{p1} as the LS problem is convex with $s + 3$ being an initial iteration for the rest of the non-malicious S-ADMM.

\label{detection_w2}
\end{algorithm}

Note that in Step 2 of Algorithm \ref{detection_w2}, the only way for a malicious PDC to fool the central PDC will be to send identical values of its dual variable over two successive iterations. However, fortunately the local PDCs do not know that the central PDC is running RR-ADMM and that too in what order. Therefore, this is a rather unlikely situation. Even if the malicious PDCs send  incorrect values of the dual variables, the detection is still possible as \eqref{diff_ww} will still hold. That is, every good PDC must obey Step 2 for the detection to be successful.

\begin{figure}
\centering
\includegraphics[width=10.4cm,height=8.5cm]{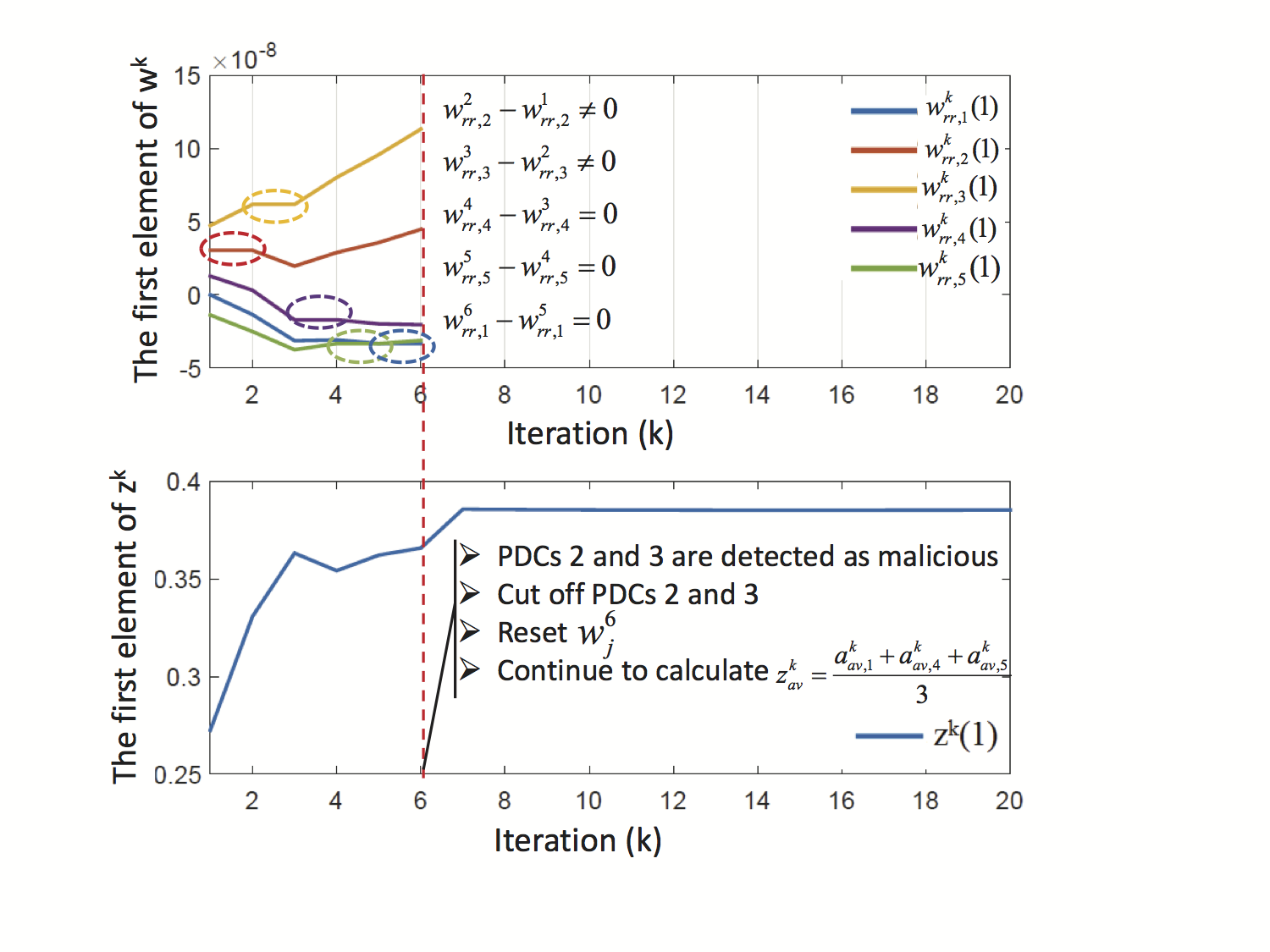}
\caption{The response of ${\boldsymbol z}^k$ before and after detection of attack with small biases using RR-ADMM .}
\label{rr_det_s_est_after}
\end{figure}

Consider the same example as in Fig. \ref{s_dec_est_after}. Fig. \ref{rr_det_s_est_after} shows the trajectory of the first element of the vectors ${\boldsymbol z}^k$ pre and post detection. In the top figure of Fig. \ref{rr_det_s_est_after},
${\boldsymbol w}_{rr, 2}^2(1) - {\boldsymbol w}_{rr, 2}^1(1) = 10^{-10}$ and ${\boldsymbol w}_{rr, 3}^3(1) - {\boldsymbol w}_{rr, 3}^2(1) = 1.2\times10^{-10}$. The central PDC also calculates ${\boldsymbol w}_{rr, 4}^4 - {\boldsymbol w}_{rr, 4}^3 = {\boldsymbol 0}_{2n \times 1}$, ${\boldsymbol w}_{rr, 5}^5 - {\boldsymbol w}_{rr, 5}^4 = {\boldsymbol 0}_{2n \times 1}$, and ${\boldsymbol w}_{rr, 1}^6 - {\boldsymbol w}_{rr, 1}^5 = {\boldsymbol 0}_{2n \times 1}$, which means PDCs 1, 4, and 5 are non-malicious. After iteration 6, the central PDC cuts off communication with the malicious PDCs, and calculates the consensus vector using S-ADMM which is the same as the one in Fig. \ref{s_dec_est_after}. The bottom figure of Fig. \ref{rr_det_s_est_after} shows the consensus variable ${\boldsymbol z}_{rr}^k$ before detection and ${\boldsymbol z}_{av}^k$ after detection.
Notice that at iteration $7$, ${\boldsymbol a}_j^7$ is reinitialized to the value based on ${\boldsymbol z}_{rr}^6$, and the dual variable ${\boldsymbol w}_j^7$ is reset to its initial value.

For calculating the differences of the dual variables over two successive iterations of RR-ADMM, the central PDC has to process only one number at every iteration; while from the S-ADMM algorithm (\ref{p3}) it follows that the central PDC has to process at least $\frac{N(N-1)}{2}$ numbers of computations at every iteration. Thus if $N$ is large then RR-ADMM will require much less computation time than S-ADMM.
Table \ref{suma1} compares S-ADMM and RR-ADMM for detection of small biases.

\begin{table*}[h!]
\centering
\caption{S-ADMM and RR-ADMM for detection with multiple small biases}
\begin{tabular}{c|c| c |c}
\hline
& Variable used for detection & Minimum \# of computations per iteration & Minimum \# of iterations needed to identify  \\
\hline
S-ADMM & ${\boldsymbol a}_{av, j}^k$ & $\frac{N(N-1)}{2}$ & 1  \\
\hline
RR-ADMM & ${\boldsymbol w}_{rr, j}^k$ & $1$ & $N+1$ \\
\hline
\end{tabular}
\label{suma1}
\end{table*}

\section{Comparison between S-ADMM and RR-ADMM}

It must be noted, again, the RR-ADMM should only be used for catching data-manipulators, not to solve the optimal solution. In fact, in this section we show that the steady-state solutions of S-ADMM and RR-ADMM with $\Delta_j^k=0$ are different from each other, which means that if the central PDC only executes RR-ADMM it will not be able to obtain an accurate solution. From Eq. \eqref{r3_2}, the value of ${\boldsymbol z}_{rr}^k$ depends on the coefficient $\alpha$. Generally, if $\alpha \ne 1$, ${\boldsymbol z}_{rr}^k \ne {\boldsymbol z}_{av}^k$ with $\Delta_j^k=0$. Thus, we only consider the case of $\alpha = 1$.

We assume ${\boldsymbol w}_{av, j}^0 = {\boldsymbol w}_{rr, j}^0$ and ${\boldsymbol z}_{av}^0 = {\boldsymbol z}_{rr}^0$. Thus, ${\boldsymbol a}_{av, j}^1 = {\boldsymbol a}_{rr, j}^1$, $1 \le j \le N$. However, ${\boldsymbol z}_{av, j}^1 \ne {\boldsymbol z}_{rr, j}^1$, because in each iteration
${\hat{\boldsymbol H}}_i \ne {\hat{\boldsymbol H}}_j $ and ${\boldsymbol a}_{av, j}^1 \ne {\boldsymbol a}_{av, i}^1$, $1 \le i, j \le N$, $i \ne j$.  When $k \rightarrow \infty$, ${\boldsymbol z}_{av}^k = {\boldsymbol a}_{av, j}^k$. In that situation, without loss of generality, we assume ${\boldsymbol z}_{rr}^k = {\boldsymbol a}_{rr, j}^k$. Thus ${\boldsymbol z}_{av}^k  - {\boldsymbol z}_{rr}^k  = {\boldsymbol a}_{av, j}^k - {\boldsymbol a}_{rr, j}^k$. Let $A_j^k = (({\hat {\boldsymbol H}}_j^k)^{\rm T}{\hat {\boldsymbol H}}_j^k + \rho {\boldsymbol I}_{2n})^{-1}$, and $B_j^k = ({\hat {\boldsymbol H}}_j^k)^{\rm T}{\boldsymbol c}_j^k - {\boldsymbol w}_j^0$. According to Eqs. \eqref{p3_2} and \eqref{r3_2}, when $k\rightarrow\infty$, the difference between ${\boldsymbol a}_{av, j}^{k+1}$ and ${\boldsymbol a}_{rr, j}^{k+1} $ can be deduced as
\begin{align}\label{differenceav_rr}
&{\boldsymbol a}_{av, j}^{k+1} - {\boldsymbol a}_{rr, j}^{k+1} = \rho \sum\limits_{i = 1}^k f^i (A_j^1, \ldots, A_j^k,\rho)({\boldsymbol z}_{av}^i - {\boldsymbol z}_{rr}^i ),
\end{align}
where
\begin{align}
&f^1(A_j^1, \ldots, A_j^k,\rho)= \nonumber\\
&(-1)^{(k-1)}\prod\limits_{i = 1}^k A_j^i \rho^{k-1} +\sum\limits_{l = 0}^{k-1}(-1)^l \prod\limits_{i = k-l}^k A_j^i \rho^l
\nonumber\\
+&(-1)^{k-2}\sum\limits_{l = 2}^{k-1}\frac{2\rho^{k-2}\prod\limits_{i = 1}^k A_j^i}{A_j^l} + \ldots + (-1)\sum\limits_{l = 1}^{k-1}\rho A_j^k A_j^l, k\ge2,
\end{align}
\begin{align}
&f^2(A_j^2, \ldots, A_j^k,\rho)
=\nonumber\\
&(-1)^{(k-2)}2\prod\limits_{i = 2}^k A_j^i \rho^{k-2}
+\sum\limits_{l = 0}^{k-2}(-1)^l \prod\limits_{i = k-l}^k A_j^i \rho^l
\nonumber\\
+&(-1)^{k-3}\sum\limits_{l = 3}^{k-1}\frac{2\rho^{k-3}\prod\limits_{i = 2}^k A_j^i}{A_j^l} + \ldots + (-1)\sum\limits_{l = 2}^{k-1}\rho A_j^k A_j^l, k\ge3,
\end{align}
\begin{align}
&\vdots\nonumber\\
&f^k(A_j^k) = 2 A_j^k, k \ge 1.
\end{align}
The expressions for the various functions in  \eqref{differenceav_rr} are derived in the Appendix \ref{appendix 2}.

Since ${\boldsymbol z}_{av}^k \ne {\boldsymbol z}_{rr}^k$, $k \ge 1$, and the value of $f^i(A_j^1, \ldots, A_j^k,\rho)$, $1 \le i \le k$, are not equal to zero in general, the difference
between ${\boldsymbol a}_{av, j}^k$ and ${\boldsymbol a}_{rr, j}^k $ is not equal to zero.

In fact, the following corollary can be written from \eqref{differenceav_rr}.

{\it Corollary I:} ${\boldsymbol z}_{rr}^k= {\boldsymbol z}_{av}^k$, $k \rightarrow \infty$ only if ${\hat{\boldsymbol H}}_i = {\hat{\boldsymbol H}}_j $, ${\boldsymbol a}_{av, i}^0 = {\boldsymbol a}_{av,j}^0 ={\boldsymbol a}_{rr, i}^0 = {\boldsymbol a}_{rr,j}^0  $, and ${\boldsymbol w}_{av,i}^0 = {\boldsymbol w}_{av, j}^0 = {\boldsymbol w}_{rr,i}^0 = {\boldsymbol w}_{rr, j}^0$, $1 \le i, j \le N$, $ i \ne j$.

\begin{proof}
Based on Eq. \eqref{p3_2}, because ${\hat{\boldsymbol H}}_i = {\hat{\boldsymbol H}}_j$ and ${\boldsymbol a}_{av, i}^0 = {\boldsymbol a}_{av,j}^0$, $1 \le i, j \le N$, $ i \ne j$,
${\boldsymbol a}_{av, i}^1 = {\boldsymbol a}_{av,j}^1 = {\boldsymbol z}_{av}^1$.
Then this result could be extended to any iteration, i.e.,
${\boldsymbol a}_{av, i}^k = {\boldsymbol a}_{av,j}^k = {\boldsymbol z}_{av}^k$. Due to the same update equations of ${\boldsymbol a}_{av}^k$ and ${\boldsymbol a}_{rr}^k$, and the same initial values, ${\boldsymbol a}_{av, i}^k ={\boldsymbol a}_{rr, i}^k ={\boldsymbol z}_{av}^k ={\boldsymbol z}_{rr}^k $ for every $k$, which follows directly from \eqref{differenceav_rr}.
\end{proof}

\section{Simulation results}\label{simulation}

\begin{figure*}
\centering
\includegraphics[width=5in]{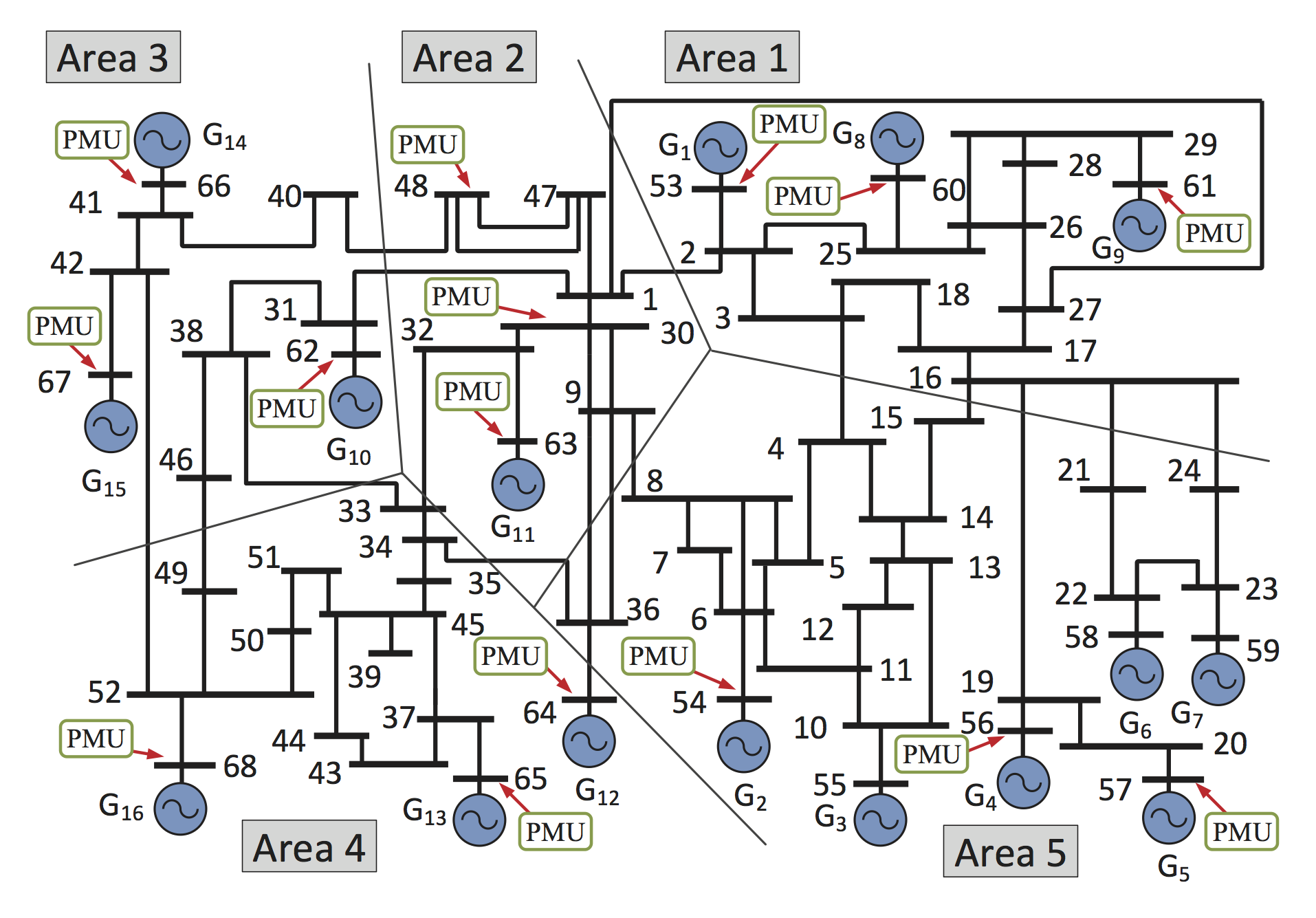}
\caption{IEEE 68-bus model}
\rule[1ex]{7in}{0.5pt}
\label{IEEE68}
\end{figure*}

\begin{figure}
\centering
\includegraphics[width=8.6cm,height=6.5cm]{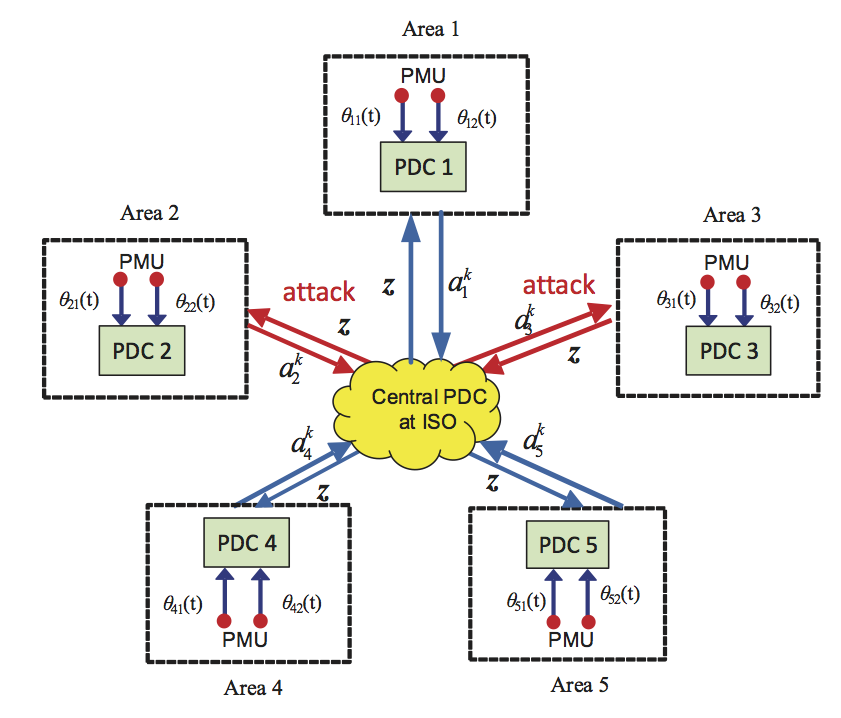}
\caption{Architecture for a 5-area power system network with 2 malicious PDCs.}
\label{D11}
\end{figure}

To verify our algorithms we consider the IEEE 68-bus power system model shown in Fig. \ref{IEEE68}. The system is divided into 5 areas, each with one local PDC and 3 PMUs. Fig. \ref{D11} shows the communication between the 5 local PDCs and the central PDC. The red lines denote that the PDCs sending messages through these links are attacked. In this case, PDC 2 and PDC 3 are malicious. The simulated measurements are obtained using the Power System Toolbox (PST) nonlinear dynamics simulation routine $\tt s\_simu$ and the data file $\tt data16m.m$ \cite{pst}. We set $\rho = 10^{-6}$. The synchronous generators are assumed to have $6^{th}$-order models for simplicity. Since there are $16$ generators, our proposed algorithm should ideally solve a $96^{th}$-order polynomial. However, our previous work on this model as reported in \cite{behzad2} show that choosing $2n =40$ yields a reasonably satisfactory estimate of the inter-area modes. The true values of the four inter-area modes  $-\sigma_i \pm j\Omega_i$, $1 \le i \le 4$, obtained by PST are $-0.32557 \pm j2.2262$, $-0.31429 \pm j3.2505$, $-0.43118 \pm j3.5809$, and $-0.43011\pm j4.9836$.

Next, we use the ADMM algorithm to estimate the oscillation modes. The following cases are considered for the purpose of illustrating the results.

{\it Case 1:) S-ADMM with no attack}.

We first deploy the standard ADMM algorithm with no attacks. The four inter-area modes $-\sigma_i \pm j\Omega_i$, $1 \le i \le 4$, converge to their global values after 25 iterations. These values are estimated to be $-\sigma_1 \pm j\Omega_1 = -0.3250 \pm j 2.2231$,
$-\sigma_2 \pm j\Omega_2 = -0.3144 \pm j3.2537$, $-\sigma_3 \pm j\Omega_3 = -0.4325 \pm 3.5847i$, and $-\sigma_4 \pm j\Omega_4 = -0.4296 \pm j4.9862$, all of which are close to their true values.

\begin{figure}
\centering
\includegraphics[width=8.6cm,height=6.5cm]{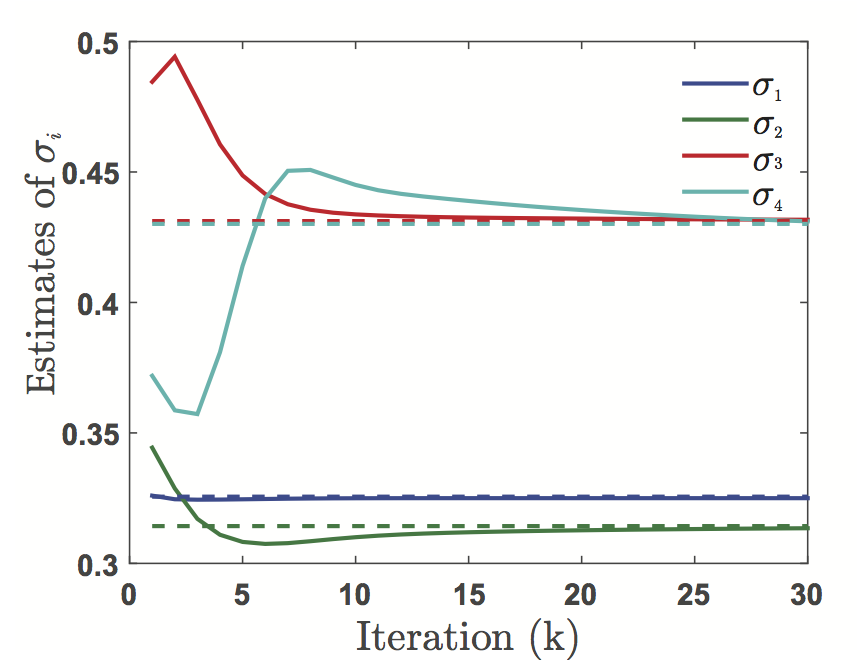}
\caption{Four estimated oscillation modes using S-ADMM with no attack.}
\label{s_ADMM_nobias}
\end{figure}

Fig. \ref{s_ADMM_nobias} shows four selected estimated modes, $\sigma$ and $\Omega$ per iteration. They converge to their global values within 25 iterations. The dashed lines show the actual values of $\sigma$ and $\Omega$ for these four modes obtained from PST.

{\it Case 2:) S-ADMM with attacks}.

\begin{figure}
\centering
\includegraphics[width=8.6cm,height=6.5cm]{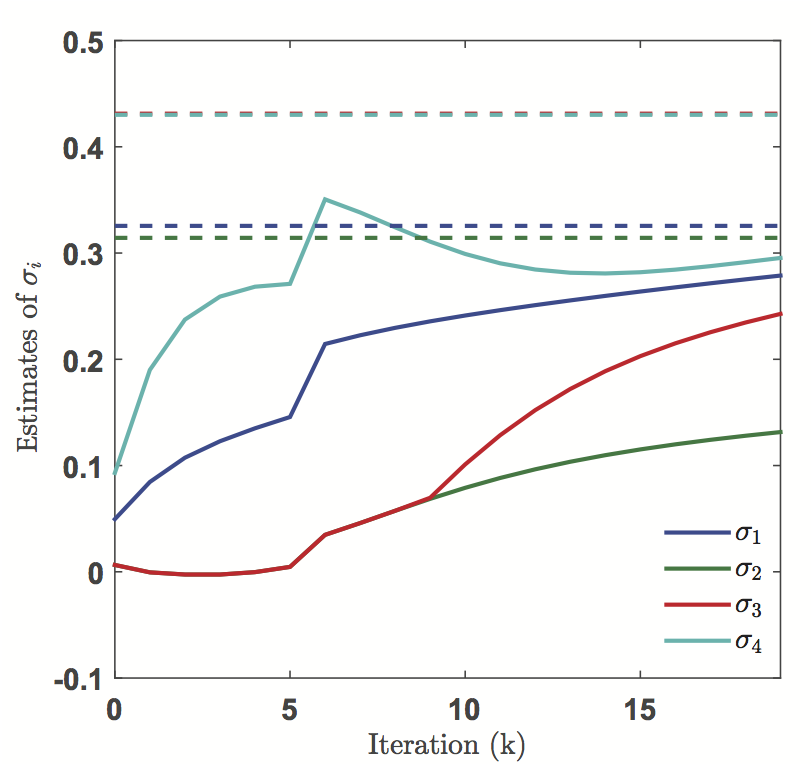}
\caption{Four estimated inter-area modes using S-ADMM with attack. The corresponding four true values of $\sigma_i$ are shown by dashed lines.}
\label{s_ADMMdis}
\end{figure}

Next we assume that PDCs in Area 2 and 3 are compromised, and respectively sends messages with arbitrary biases $\Delta_2^k$ and $\Delta_3^k$, generated using the random number generator in Matlab. This architecture is shown in Fig. \ref{D11}.
Fig. \ref{s_ADMMdis} shows the estimations of $\sigma_i$, $1 \le i \le 4$, per iteration for the four inter-area modes. The dashed lines show the actual values of $\sigma_i$. Same holds for the estimates of $\Omega$. It can be seen that without detection and mitigation, the trajectories of the estimates keep diverging due to the additive effect of the bias.

{\it Case 3:) S-ADMM for detecting data manipulators with general biases}.

We next apply Algorithm \ref{detecting_S-ADMM} to catch the malicious PDCs 2 and 3.
\begin{figure}[!t]
\centering
\includegraphics[width=8.5cm,height=6.5cm]{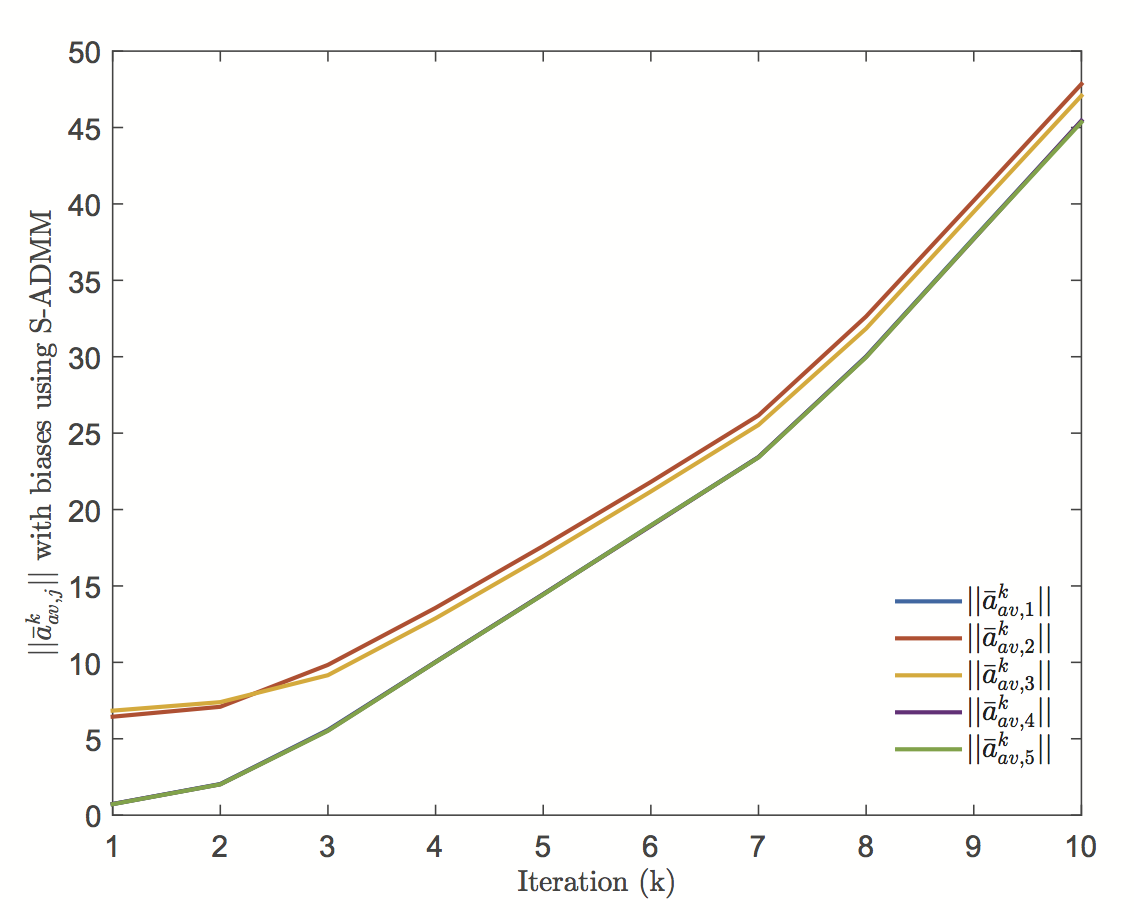}
\caption{Evolution of $||{\bar {\boldsymbol  a}}_{av, j}^k||$ when S-ADMM is run under attacks.}
\label{multi_s_rand}
\end{figure}
 Fig. \ref{multi_s_rand} shows the norms of the vector ${\bar{\boldsymbol a}}_{av, i}^{k}$, $1 \le i \le N$. At the first iteration, $||{\bar{\boldsymbol a}}_{av, 1}^1|| = 0.7286$, $||{\bar{\boldsymbol a}}_{av, 2}^1|| = 6.4435$, $||\bar{{\boldsymbol a}}_{av, 3}^1|| = 6.839$, $||{\bar{\boldsymbol a}}_{av, 4}^1|| = 0.7313$, $||{\bar{\boldsymbol a}}_{av, 5}^1|| = 0.7189$. The threshold $\gamma_a^1 = 0.0485$. ${\bar{\boldsymbol a}}^1_{av} $ is separated into three groups: ${\bar{\boldsymbol a}}_{av, 1}^1$, ${\bar{\boldsymbol a}}_{av, 4}^1$, and ${\bar{\boldsymbol a}}_{av, 5}^1$ are in the first group, ${\bar{\boldsymbol a}}_{av, 2}^1$ and ${\bar{\boldsymbol a}}_{av, 3}^1$ are in the second and third groups, respectively. According to Algorithm \ref{detecting_S-ADMM}, the minimum value $||{\bar{\boldsymbol a}}_{av, 5}^1||$ is in the first group. Therefore, PDCs 1, 4, and 5 are identified as non-malicious, which matches the true situation. Fig. \ref{s_admm_dec_mode_aft} shows the estimations of $\sigma_i$, $1 \le i \le 4$ for the four inter-area modes after PDCs 2 and 3 are disconnected. The dashed lines show the actual values of $\sigma_i$. The final values of the estimates of $\sigma_i$ in this case match with their actual values again.
\begin{figure}
\centering
\includegraphics[width=8.5cm,height=6.7cm]{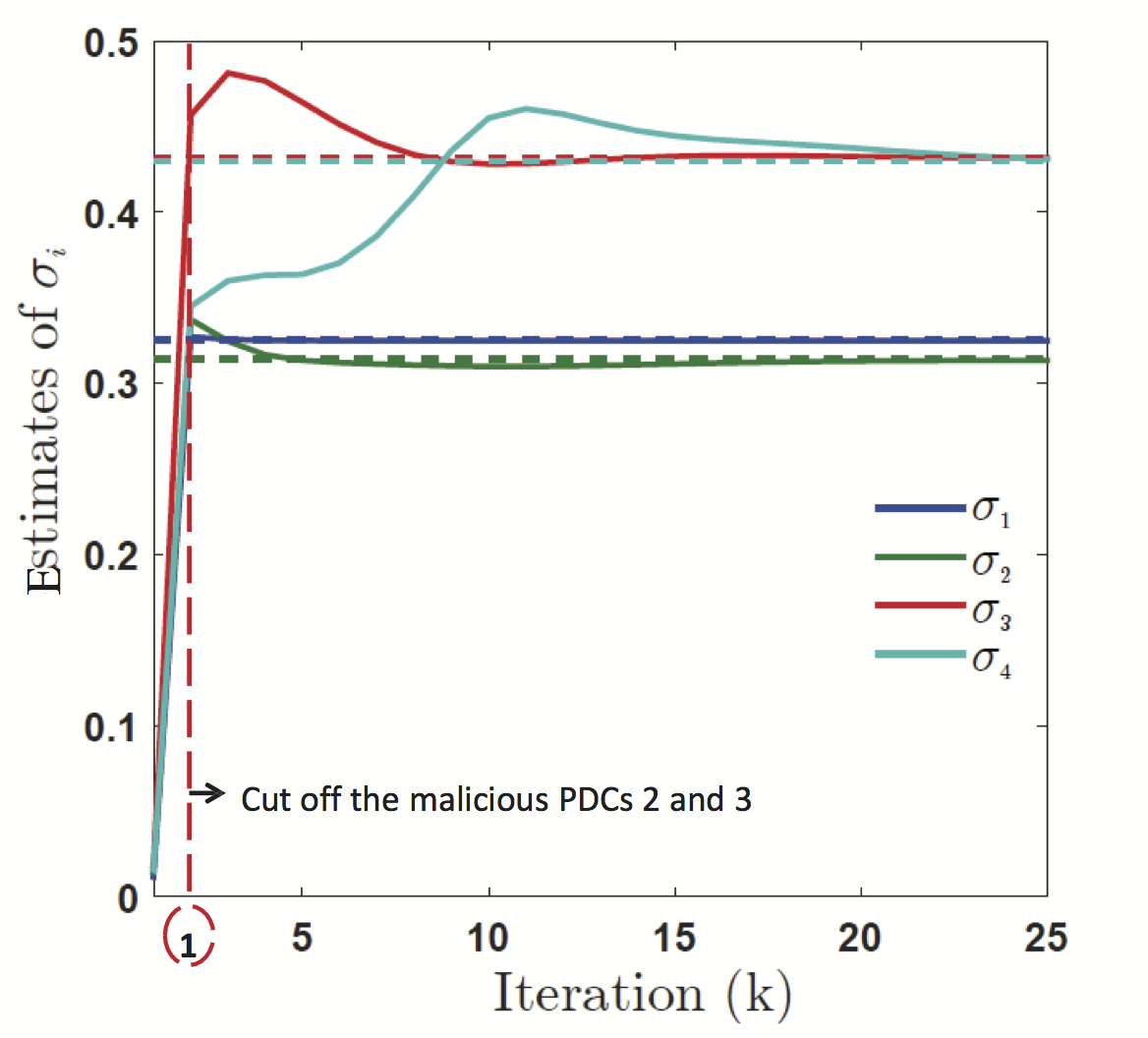}
\caption{Values of the real parts of the four inter-area modes before and after detection using S-ADMM. True values of $\sigma_i$ are shown by dashed lines.}
\label{s_admm_dec_mode_aft}
\end{figure}

\begin{figure}[!t]
\centering
\includegraphics[width=8.5cm,height=6.5cm]{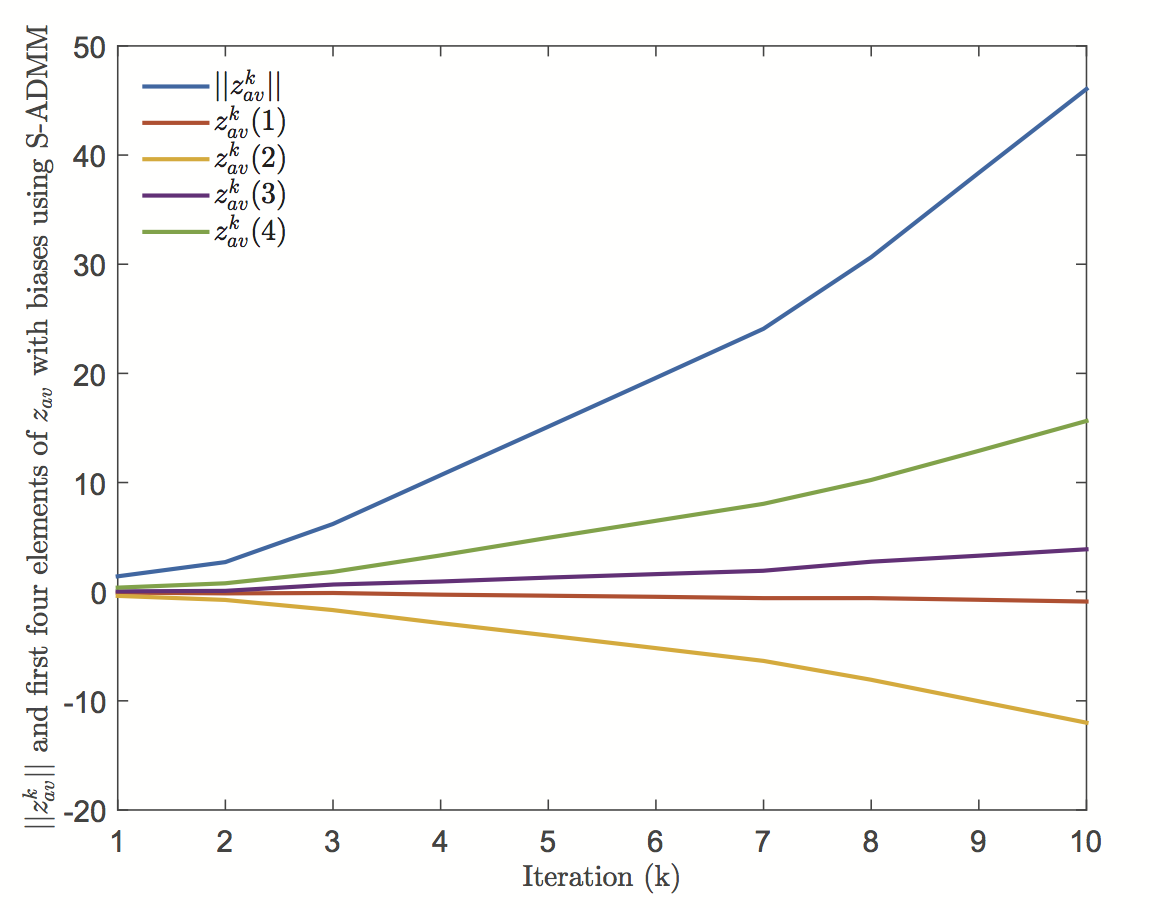}
\caption{Evolutions of $||{\boldsymbol  z}_{av}^k||$ and the first four elements of ${\boldsymbol z}_{av}^k$ when S-ADMM is run under attacks.}
\label{multi_zav_rand}
\end{figure}

Considering the same example, Fig. \ref{multi_zav_rand} shows the norm and the first four elements of the consensus vector ${\boldsymbol z}_{av}^k$ using S-ADMM.
None of these trajectories has any signature of the biases from PDCs 2 and 3, and hence it is impossible to identify them by just tracking ${\boldsymbol z}_{av}^k$.

{\it Case 4:) RR-ADMM for detecting data manipulators with general biases}.

\begin{figure}[!t]
\centering
\includegraphics[width=8.5cm,height=6.5cm]{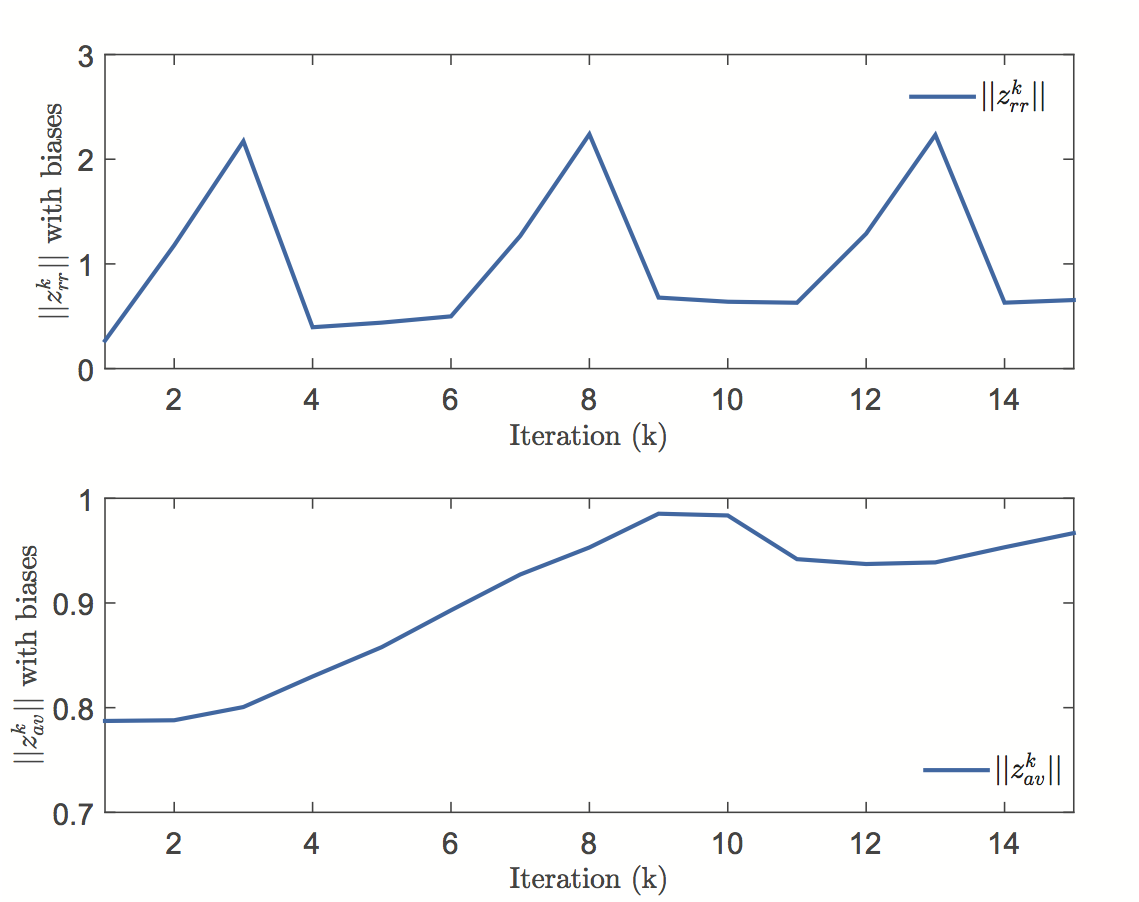}
\caption{Evolutions of $||{\boldsymbol  z}^k||$ when RR-ADMM and S-ADMM are run under attacks.}
\label{multi_rr_z_rand_detc}
\end{figure}
\begin{figure}
\centering
\includegraphics[width=8.5cm,height=6.7cm]{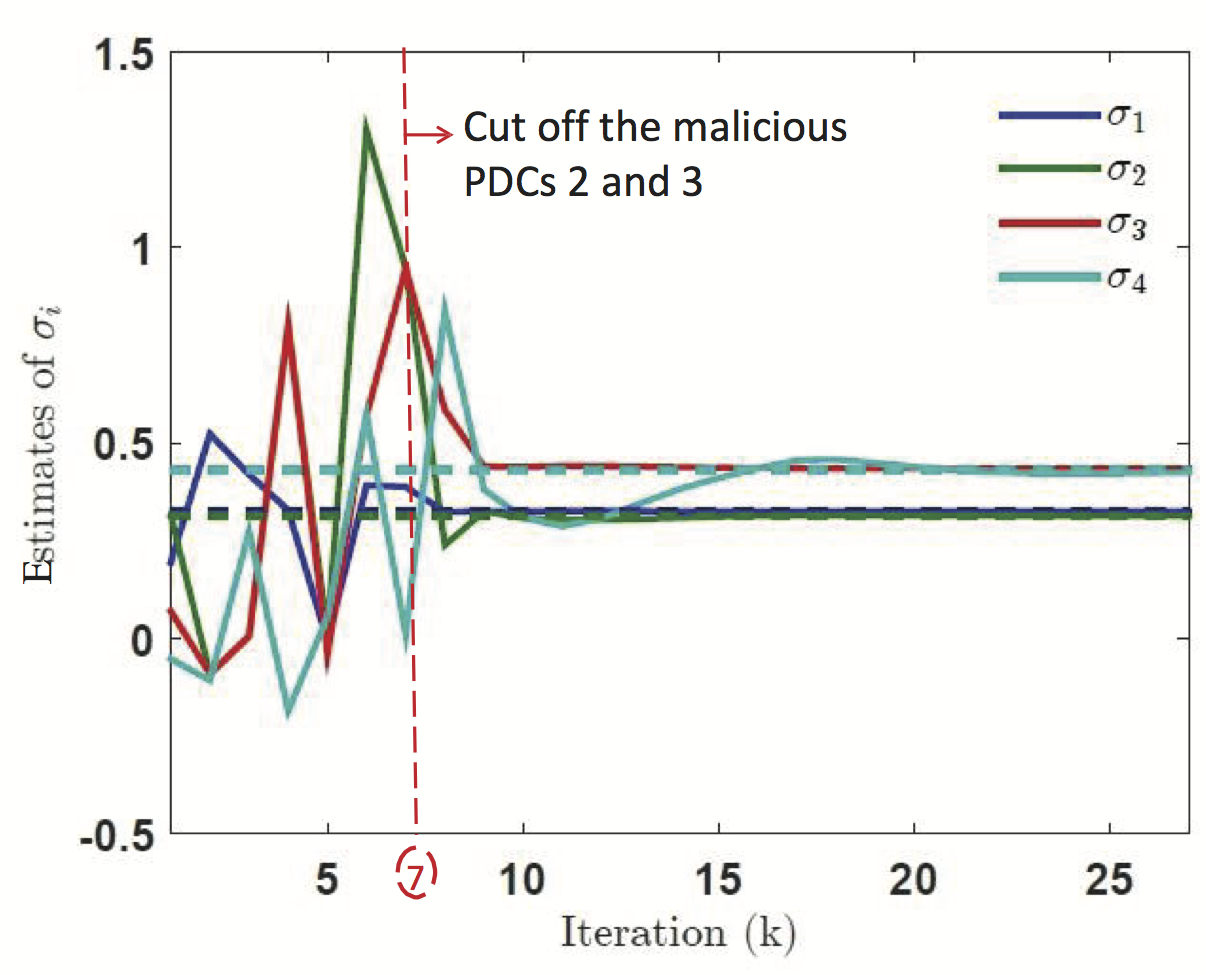}
\caption{Values of the real parts of the four inter-area modes before and after detection with RR-ADMM. True values of $\sigma_i$ are shown by dashed lines.}
\label{rr_admm_dec_suc}
\end{figure}

We next apply Algorithm 2 to detect the malicious PDCs 2 and 3.
The top figure of Fig. \ref{multi_rr_z_rand_detc} depicts the norm of the consensus vector ${\boldsymbol z}_{rr}^k$ using RR-ADMM. From iteration 1 to iteration 5, $||{\boldsymbol z}_{rr}^1|| = 0.267$ is minimum. $||{\boldsymbol z}_{rr}^6|| = 0.4991$ and the threshold is $\gamma_z = 0.2321$.  $||{\boldsymbol z}_{rr}^2||>||{\boldsymbol z}_{rr}^1||+ \gamma_z$ and $||{\boldsymbol z}_{rr}^3||>||{\boldsymbol z}_{rr}^1|| + \gamma_z$, so PDCs 2 and 3 are identified as malicious.
The bottom panel of Fig. \ref{multi_rr_z_rand_detc} shows the norm of the consensus vector ${\boldsymbol z}_{av}^k$ using S-ADMM. The trajectory of $||{\boldsymbol z}_{av}^k||$ does not contain any attack signature, and hence the malicious PDCs cannot be identified from it. Fig. \ref{rr_admm_dec_suc} shows the estimations of $\sigma_i$, $1 \le i \le 4$ for the four inter-area modes after PDCs 2 and 3 are disconnected using RR-ADMM. The dashed lines show the actual values of $\sigma_i$. The final values of
the estimates of $\sigma_i$ in this case match with their true values again.

{\it Case 5:) RR-ADMM for detecting the data manipulators with random order}.

\begin{figure}
\centering
\includegraphics[width=8.5cm,height=6.5cm]{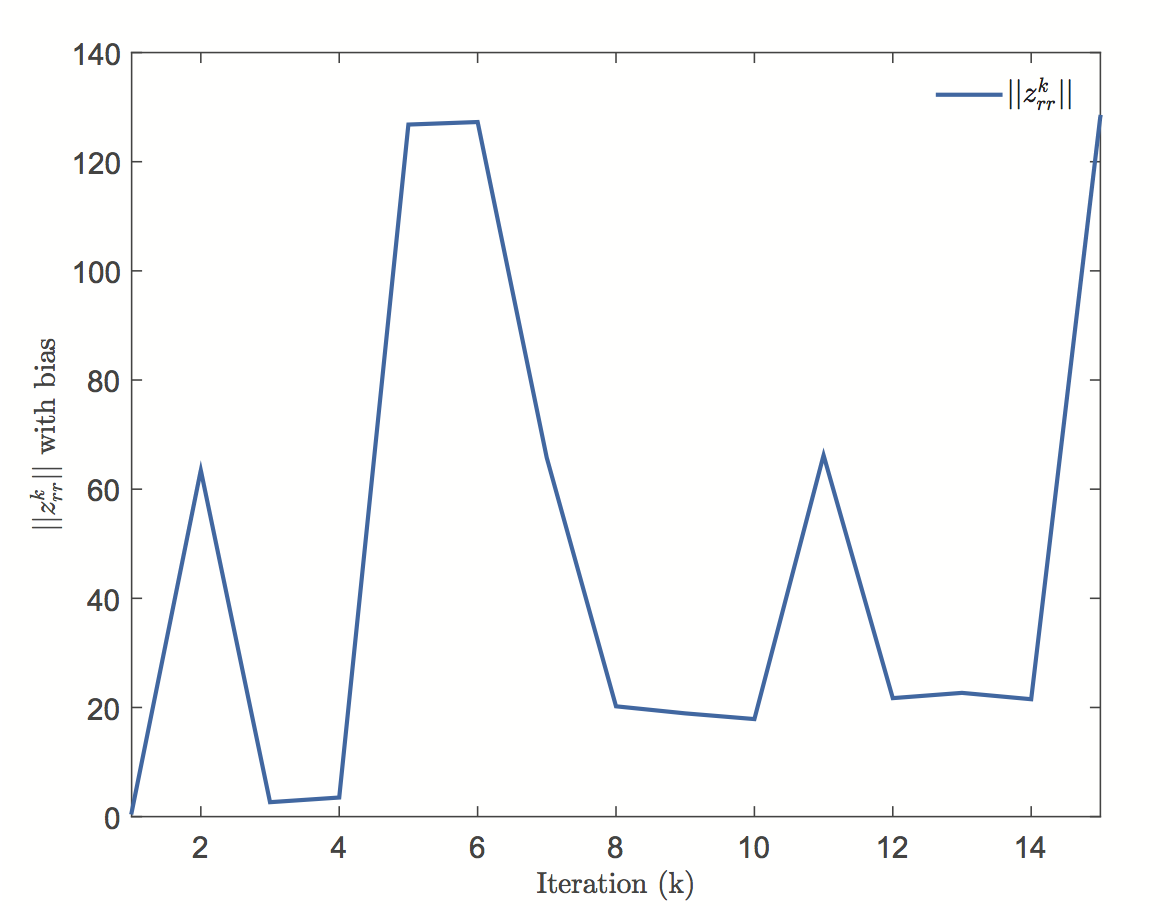}
\caption{$||{\boldsymbol z}_{rr}^k||$ with random order of RR-ADMM.}
\label{RR_DOS}
\end{figure}

In this case, the central PDC chooses the order of the local PDCs are (1, 2, 4, 5, 3) for the first period, (3, 2, 5, 4, 1) in the second period, and (2, 5, 4, 1, 3) in the third period. $\alpha = 0.9$. Fig. \ref{RR_DOS} shows the trajectory of $||{\boldsymbol z}_{rr}^k||$.
$||{\boldsymbol z}_{rr}^1|| = 0.767$, $||{\boldsymbol z}_{rr}^2|| = 63.4122$, $||{\boldsymbol z}_{rr}^3|| = 2.6447$, $||{\boldsymbol z}_{rr}^4|| = 3.5022$, and $||{\boldsymbol z}_{rr}^5|| = 126.8068$. $||{\boldsymbol z}_{rr}^1||$ is smaller than others in the first period, and $\alpha {\boldsymbol z}_{rr}^1 $ is equal to the data from the third local PDC. So in the second period, the central PDC assigns the estimate of the third PDC as the consensus variable, which is ${\boldsymbol z}_{rr}^{10}$. The threshold is $\gamma_z = ||{\boldsymbol z}_{rr}^{10}|| - ||{\boldsymbol z}_{rr}^1|| = 17.1055$. $||{\boldsymbol z}_{rr}^2||>||{\boldsymbol z}_{rr}^1|| + \gamma_z$ and $||{\boldsymbol z}_{rr}^5||>||{\boldsymbol z}_{rr}^1|| + \gamma_z$, and ${\boldsymbol z}_{rr}^2 = \alpha {\bar{\boldsymbol a}}_{rr,2}^2$ and ${\boldsymbol z}_{rr}^5 =\alpha {\bar{\boldsymbol a}}_{rr,3}^5$. Hence, the PDCs 2 and 3 are identified as malicious, which matches the true situation.

{\it Case 6:) S- and RR-ADMM for detecting data manipulators with sparse and non-sparse biases.}

\begin{figure}
\centering
\includegraphics[width=8.5cm,height=6.5cm]{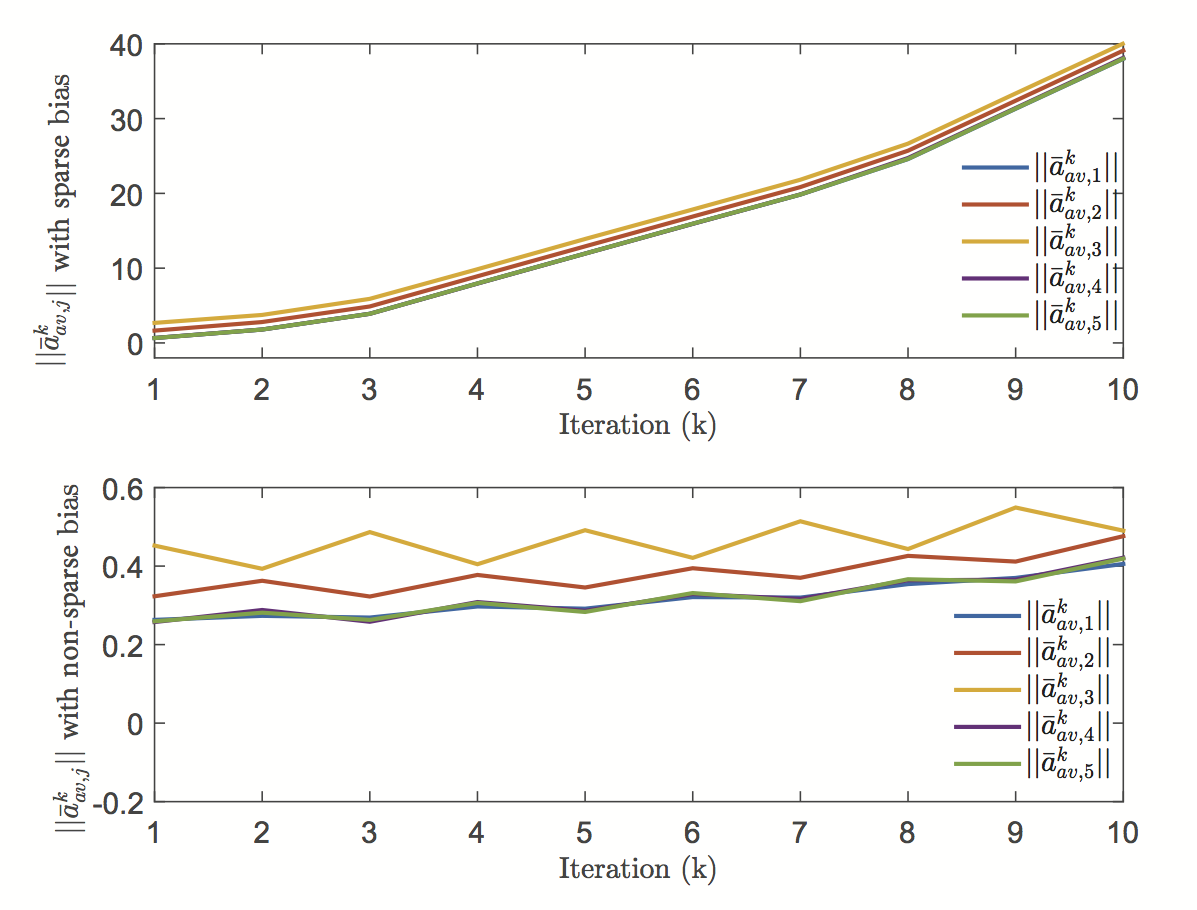}
\caption{ The response of $||{\bar{\boldsymbol a}}_{av, j}^k||$ with sparse biases and non-sparse biases using S-ADMM.}
\label{S_ADMM_def_bias}
\end{figure}
\begin{figure}
\centering
\includegraphics[width=8.5cm,height=6.5cm]{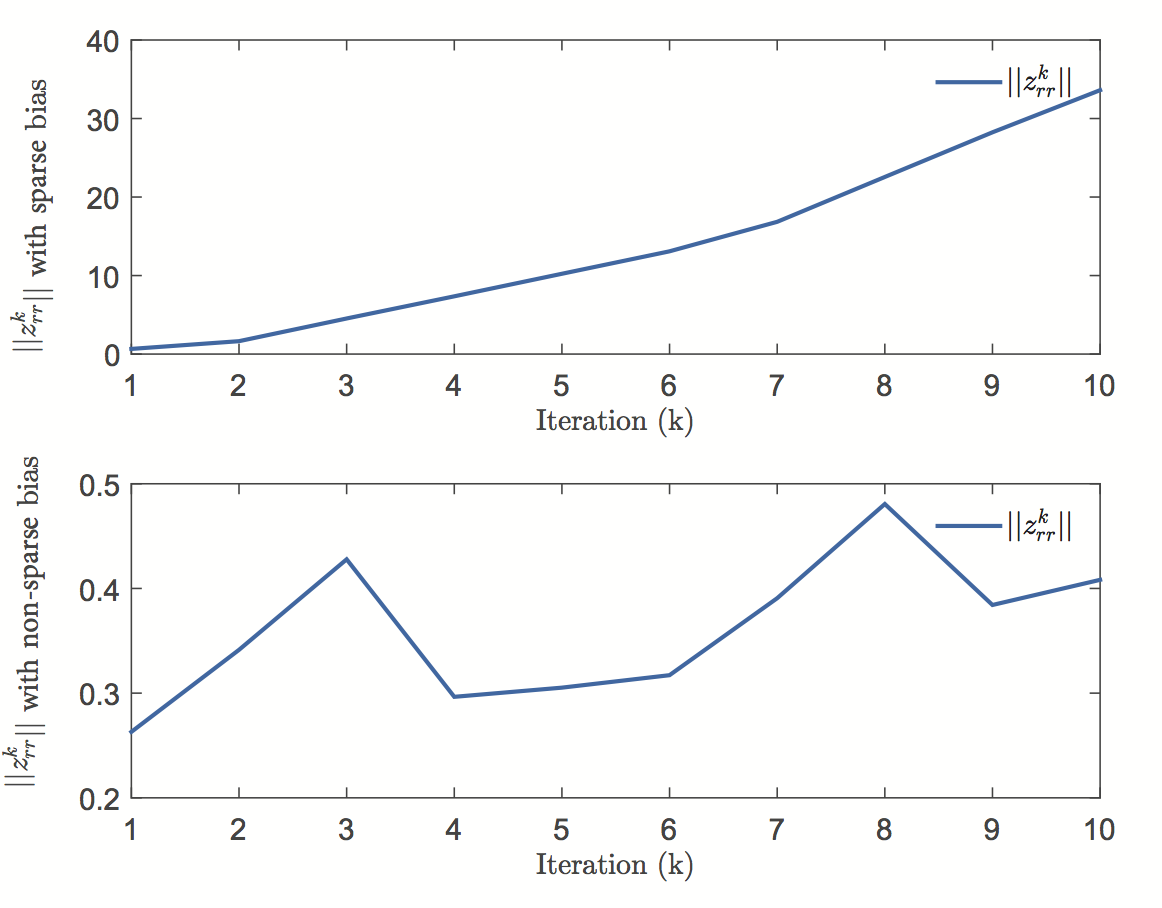}
\caption{ The response of $||{{\boldsymbol z}}_{rr}^k||$ with sparse biases and non-sparse biases using RR-ADMM.}
\label{RR_ADMM_def_bias}
\end{figure}

In this case, we test the sensitivity of S-ADMM and RR-ADMM with respect to the sparsity in the elements of the bias. We compare two scenarios. In both scenarios the biases chosen to be constant over time. We assume PDCs 2 and 3 are attacked. In the first scenario, all entries of $\Delta_2$ and $\Delta_3$ are zero except $\Delta_2(5)=0.1$ and $\Delta_3(5)=0.2$. In the second example, $\Delta_2 = \frac{0.1}{\sqrt{2n}}{\boldsymbol 1}_{2n \times 1}$ and $\Delta_3 = \frac{0.2}{\sqrt{2n}}{\boldsymbol 1}_{2n \times 1}$. Thus the norms of the respective bias vectors in the two examples are the same.
Fig. \ref{S_ADMM_def_bias} draws the trajectories of $||{\bar{\boldsymbol a}}_{av, j}^k||$, $1 \le k \le 5$, using S-ADMM in these two examples.
The top figure shows the first scenario, i.e., the case when the bias vector is sparse, while the bottom figure shows the case for non-sparse bias. For both examples, S-ADMM can catch the attacked PDCs 2 and 3 successfully by separating the estimates into three correct groups, namely PDCs (1, 4, 5) in one group and PDCs (2, 3) in other two groups, respectively.
Fig. \ref{RR_ADMM_def_bias}, on the other hand, shows the trajectories of $||{\boldsymbol z}^k_{rr}||$ using RR-ADMM for these two examples. The top figures shows the scenario with sparse bias. In the first period, $||{{\boldsymbol z}}_{rr}^1|| = 0.6579$, $||{{\boldsymbol z}}_{rr}^2|| = 1.6434$, $||{{\boldsymbol z}}_{rr}^3|| = 4.534$, $||{{\boldsymbol z}}_{rr}^4|| = 7.3559$, and $||{{\boldsymbol z}}_{rr}^5|| = 10.2282$. $||{{\boldsymbol z}}_{rr}^1||$ is minimum, and $||{{\boldsymbol z}}_{rr}^6|| = 13.0892$. Hence, the threshold is $\gamma_z = ||{{\boldsymbol z}}_{rr}^6|| - ||{{\boldsymbol z}}_{rr}^1|| = 12.4313$. $||{{\boldsymbol z}}_{rr}^k||< \gamma_z + ||{{\boldsymbol z}}_{rr}^1||$, $1 \le k \le 5$. Therefore, according to Algorithm \ref{detecting}, none of the PDCs are detected to be malicious. When the bias is non-sparse, however, RR-ADMM identifies the malicious PDCs successfully, as shown in the bottom figure. This example shows that S-ADMM is fairly insensitive to the sparsity pattern in the elements of the bias vector since it compares the estimate of every individual local PDC. RR-ADMM, on the other hand, may show false positives when the attack vector is sparse since it relies only on the consensus vector for detection.


\begin{figure}
\centering
\includegraphics[width=8.5cm,height=6.5cm]{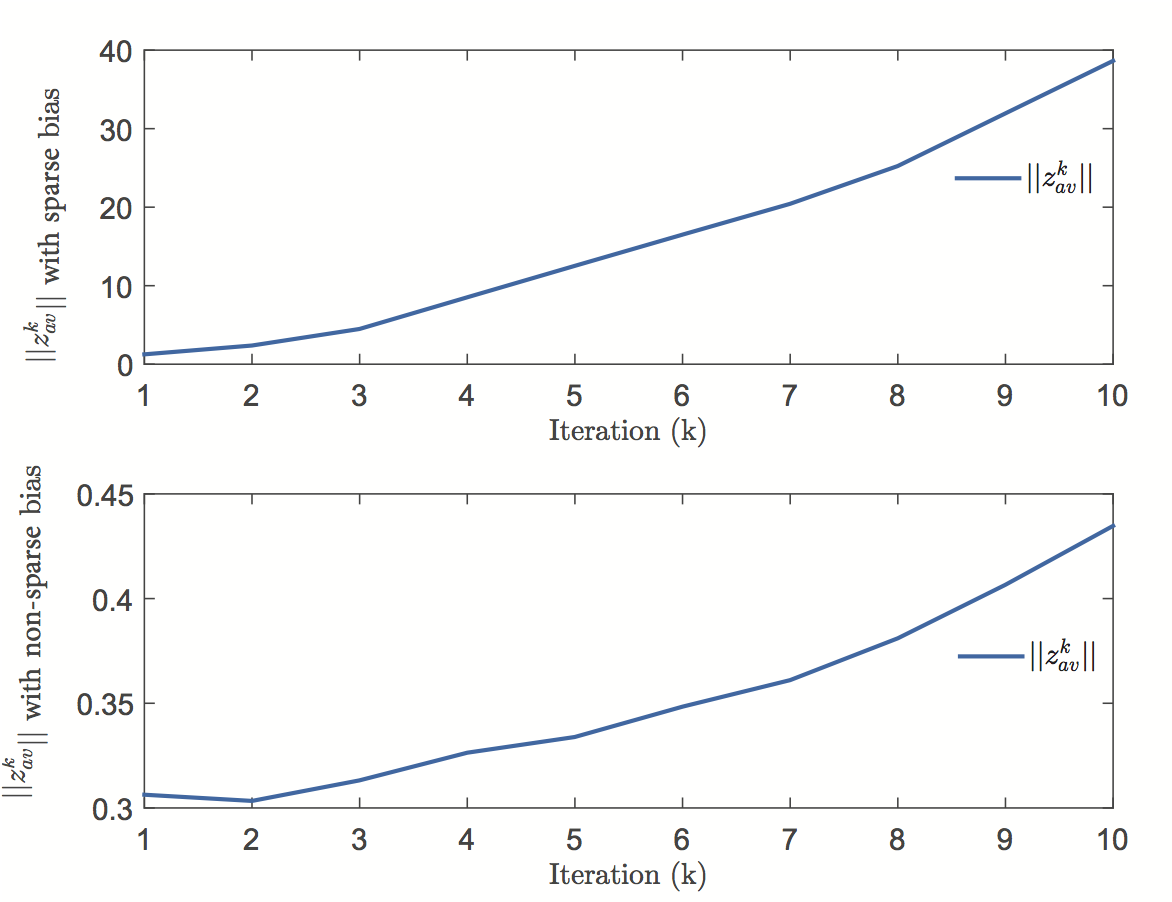}
\caption{ The response of $||{{\boldsymbol z}}_{av}^k||$ with sparse biases and non-sparse biases using S-ADMM.}
\label{S_ADMM_z_def_bias}
\end{figure}

Finally, Fig. \ref{S_ADMM_z_def_bias} shows the trajectories of $||{\boldsymbol z}^k_{av}||$ computed using S-ADMM for both sparse and non-sparse cases. Neither of these trajectories has any signature by which PDCs 2 and 3 can be identified to be malicious. This, again, clearly shows that if the central PDC wants to track only the consensus variable for detection, then S-ADMM is of no use, and it must resort to RR-ADMM.

{\it Case 7:) S-ADMM for detecting data manipulators with small biases}.

Let the second and the third PDCs be respectively injected with
$\Delta_2^k = 0.002 \times {\boldsymbol 1}_{2n \times 1}$ and $\Delta_3^k = 0.003 \times {\boldsymbol 1}_{2n \times 1}$.
\begin{figure}[!t]
\centering
\includegraphics[width=8.5cm,height=6.5cm]{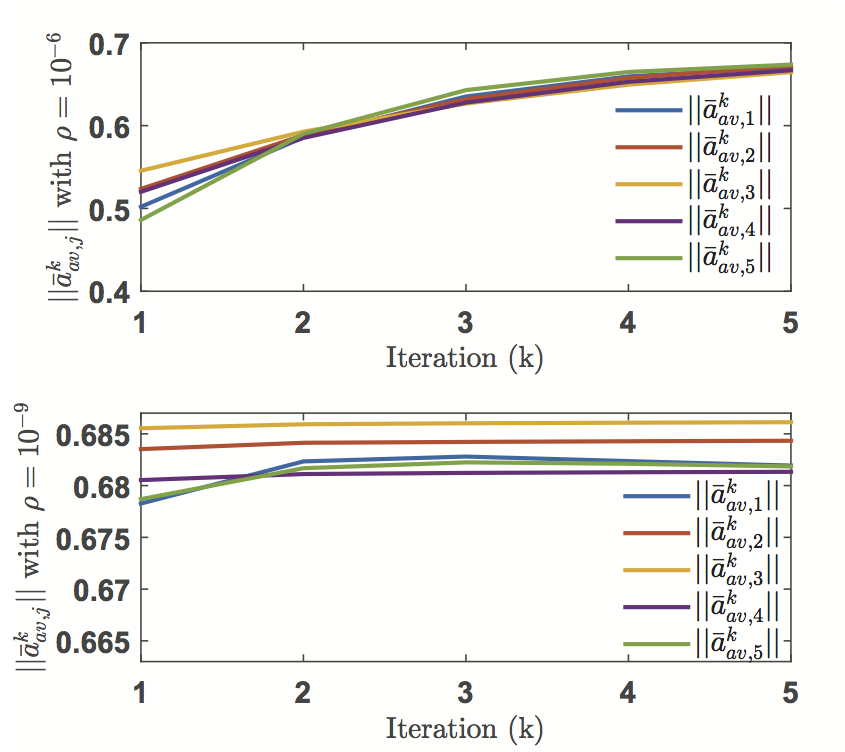}
\caption{Evolutions of $||{\bar {\boldsymbol  a}}_{av, j}^k||$ with multiple biases and different values of $\rho$ using S-ADMM. }
\label{s_admm}
\end{figure}
Fig. \ref{s_admm} shows the norm of the vector ${\bar{\boldsymbol a}}_{av, i}^{k}$, $1 \le i \le N$ with different values of $\rho$. In the top figure of Fig. \ref{s_admm}, with $\rho = 10^{-6}$ as before, at the second iteration, $||{\bar{\boldsymbol a}}_{av, 1}^2|| = 0.5857$, $||{\bar{\boldsymbol a}}_{av, 2}^2|| = 0.5884$, $||{\bar{\boldsymbol a}}_{av, 3}^2|| = 0.5927$, $||{\bar{\boldsymbol a}}_{av, 4}^2|| = 0.5852$, $||{\bar{\boldsymbol a}}_{av, 5}^2|| = 0.5902$. The threshold $\gamma_a^2 = 0.0015$. ${\bar{\boldsymbol a}}_{av}^2 $ is separated into four groups: ${\bar{\boldsymbol a}}_{av, 1}^2$ and ${\bar{\boldsymbol a}}_{av, 4}^2$ are in the first group, ${\bar{\boldsymbol a}}_{av, 2}^2$, ${\bar{\boldsymbol a}}_{av, 3}^2$, and ${\bar{\boldsymbol a}}_{av, 5}^2$ are in the second, third and fourth groups, respectively. The minimum value ${\bar{\boldsymbol a}}_{av, 4}^2$ is in the first group. Therefore, PDCs 2, 3, and 5 are treated as malicious which is a false positive. The bottom figure of Fig. \ref{s_admm} depicts the norm of ${\bar{\boldsymbol a}}_{av, j}^k$ with $\rho$ reduced to $10^{-9}$. At the second iteration, $||{\bar{\boldsymbol a}}_{av, 1}^2|| = 0.6819$, $||{\bar{\boldsymbol a}}_{av, 2}^2|| = 0.6841$, $||{\bar{\boldsymbol a}}_{av, 3}^2|| = 0.6859$, $||{\bar{\boldsymbol a}}_{av, 4}^2|| = 0.6811$, $||{\bar{\boldsymbol a}}_{av, 5}^2|| = 0.6817$. The threshold $\gamma_a^2 = 0.00084$. ${\bar{\boldsymbol a}}_{av}^2 $ is separated into three groups: ${\bar{\boldsymbol a}}_{av, 1}^2$, ${\bar{\boldsymbol a}}_{av, 4}^2$, and ${\bar{\boldsymbol a}}_{av, 5}^2$ are in the first group, ${\bar{\boldsymbol a}}_{av, 2}^2$ and ${\bar{\boldsymbol a}}_{av, 3}^2$ are in the second and third groups, respectively. $||{\bar{\boldsymbol a}}_{av, 4}^2||$ in the first group is minimum. Therefore, PDCs 1, 4, and 5 are non-malicious which matches the true situation.
Top figure of Fig. \ref{s_admm_dec_small} shows the estimations of $\sigma_i$, $1 \le i \le 4$, for the four inter-area modes. The dashed lines show the actual values of $\sigma_i$. The final values of the estimates of $\sigma_i$ in this case do not match their actual values, due to the undetected bias. The estimates are divergent, but due to the specific choice of $\Delta$, the rate of divergence is slow, and, hence, not very visible.
The bottom figure of Fig. \ref{s_admm_dec_small} shows the estimations of $\sigma_i$, $1 \le i \le 4$, for the four inter-area modes. After the malicious PDCs 2 and 3 are cut off, the final values of the estimates of $\sigma_i$ match their true values again.

\begin{figure}[!t]
\centering
\includegraphics[width=8.5cm,height=6.5cm]{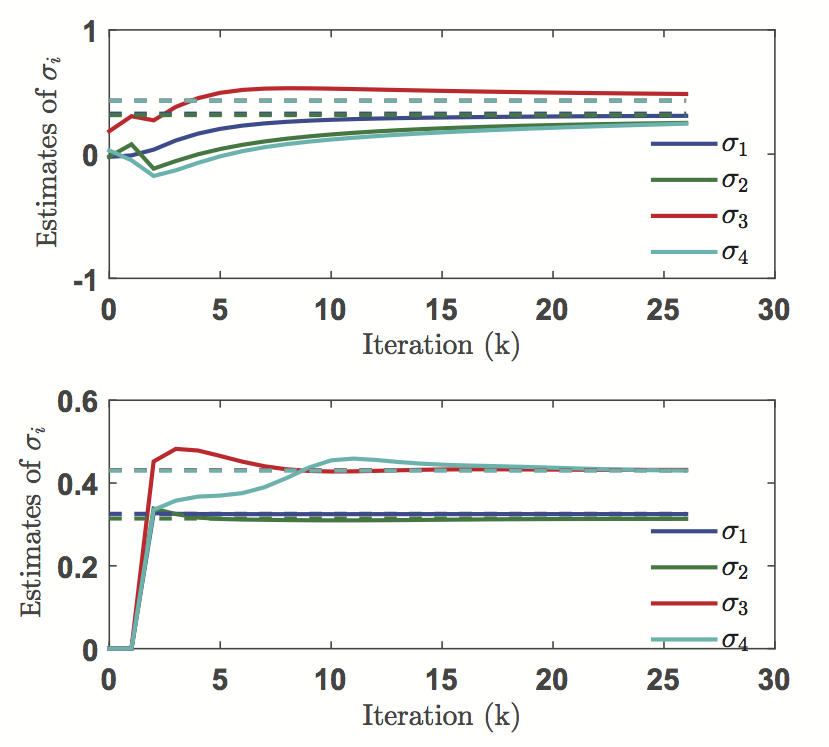}
\caption{Values of the real parts of the four estimated inter-area modes before and after detection using S-ADMM with small biases. True values of $\sigma_i$ are shown by dashed lines.}
\label{s_admm_dec_small}
\end{figure}

{\it Case 8:) RR-ADMM for detecting data manipulators with small biases}.

Consider the same example as in Case 7. 
\begin{figure}[!t]
\centering
\includegraphics[width=8.5cm,height=6.5cm]{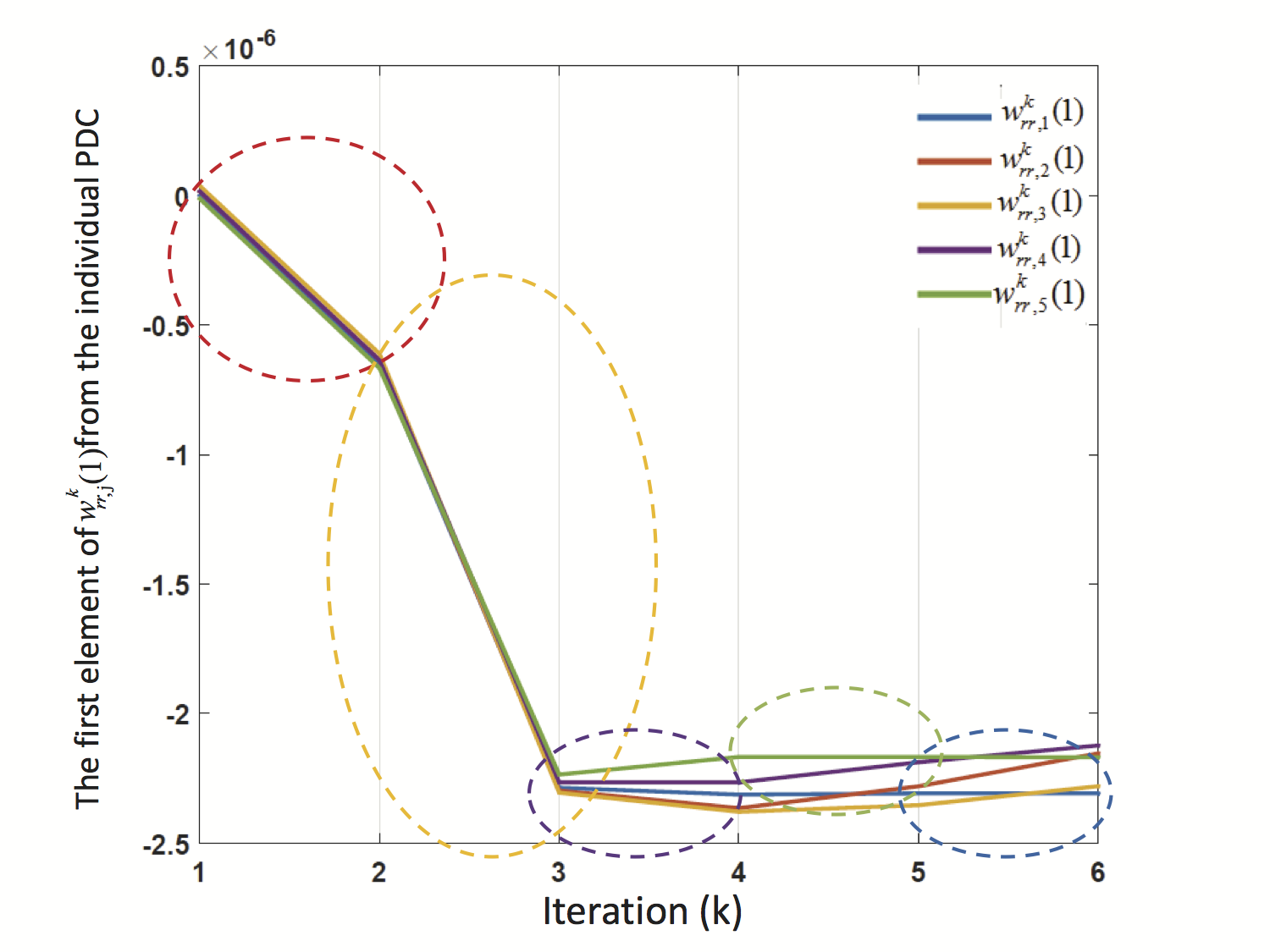}
\caption{The first element of ${\boldsymbol  w}_{rr, j}^k$ from the local PDCs.}
\label{rr_admm_w_after}
\end{figure}
Fig \ref{rr_admm_w_after} shows the first element of ${\boldsymbol  w}_{rr, j}^k$ from the individual PDCs. Only the red and yellow curves in their own color cycles are not flat, i.e., ${\boldsymbol w}_{rr, 2}^2(1) - {\boldsymbol w}_{rr, 2}^1(1) \ne 0$ and ${\boldsymbol w}_{rr, 3}^3(1) - {\boldsymbol w}_{rr, 3}^2(1)\ne 0$. The central PDC calculates ${\boldsymbol w}_{rr, 4}^4 - {\boldsymbol w}_{rr, 4}^3= {\boldsymbol 0}_{2n \times 1}$, ${\boldsymbol w}_{rr, 5}^5 - {\boldsymbol w}_{rr, 5}^4= {\boldsymbol 0}_{2n \times 1}$, and ${\boldsymbol w}_{rr, 1}^6 - {\boldsymbol w}_{rr, 1}^5= {\boldsymbol 0}_{2n \times 1}$. Thus, PDCs 2 and 3 are identified as attacked, which matches the true condition.

Consider another example: $\Delta_2^k = 0.0001 \times {\boldsymbol 1}_{2n \times 1}$ and $\Delta_3^k = 0.0002 \times {\boldsymbol 1}_{2n \times 1}$.
\begin{figure}[!t]
\centering
\includegraphics[width=8.5cm,height=6.5cm]{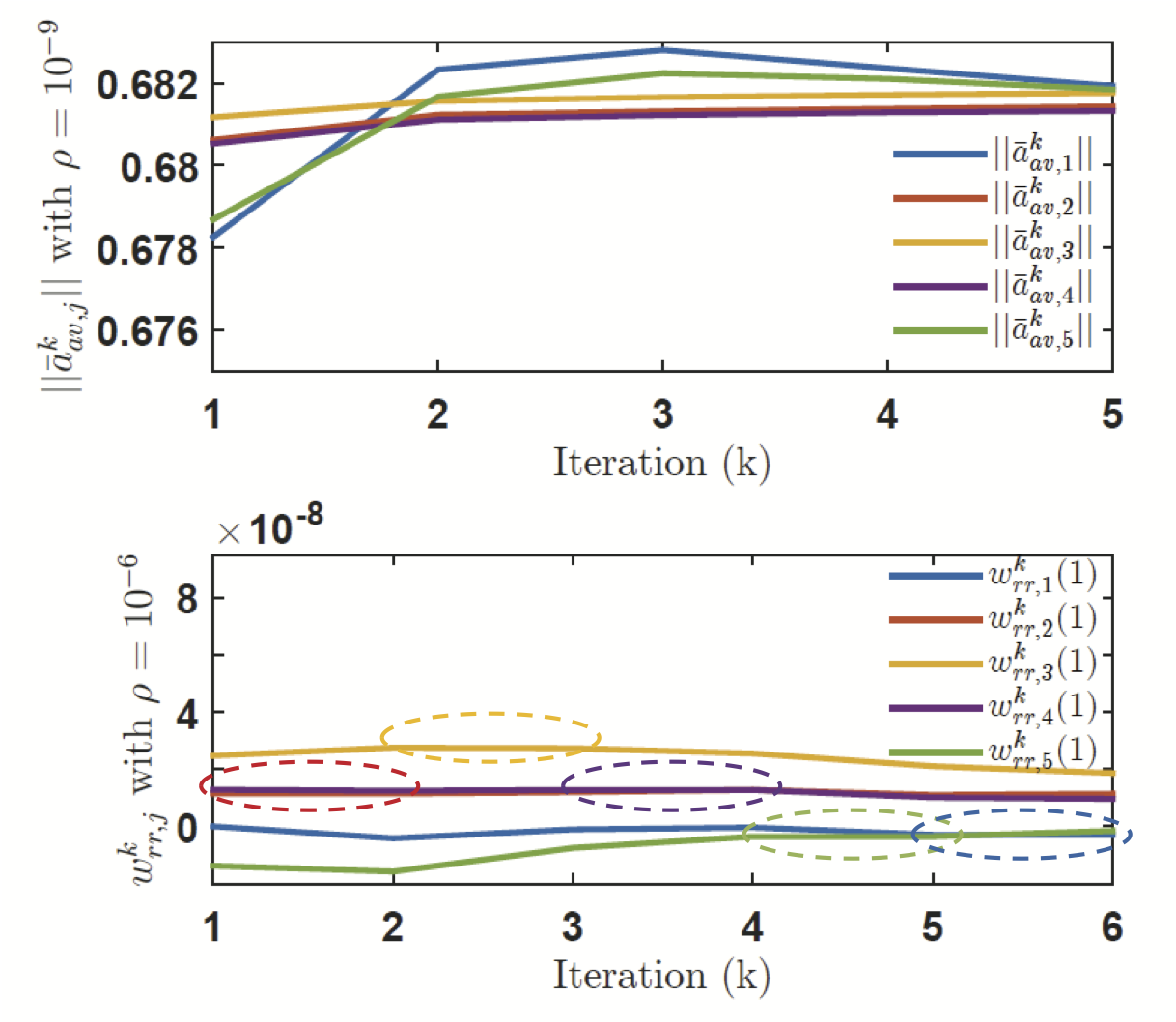}
\caption{$||{\boldsymbol  a}_{av, j}^k||$ with $\rho = 10^{-9}$ and the first element of ${\boldsymbol w}_{rr,j}^k$ with $\rho = 10^{-6}$.}
\label{s_rr_admm_after}
\end{figure}
The top figure of Fig. \ref{s_rr_admm_after} depicts the norm of ${\bar{\boldsymbol a}}_{av, j}^k$ from the individual PDCs with $\rho = 10^{-9}$. At the second iteration, $||{\bar{\boldsymbol a}}_{av, 1}^2|| = 0.6823$, $||{\bar{\boldsymbol a}}_{av, 2}^2|| = 0.6812$, $||{\bar{\boldsymbol a}}_{av, 3}^2|| = 0.6816$, $||{\bar{\boldsymbol a}}_{av, 4}^2|| = 0.6811$, $||{\bar{\boldsymbol a}}_{av, 5}^2|| = 0.6817$. The threshold $\gamma_a^2 = 0.00024$. ${\bar{\boldsymbol a}}_{av}^2 $ is separated into three groups: ${\bar{\boldsymbol a}}_{av, 2}^2$ and ${\bar{\boldsymbol a}}_{av, 4}^2$ are in the first group, ${\bar{\boldsymbol a}}_{av, 3}^2$ and ${\bar{\boldsymbol a}}_{av, 5}^2$ are in second group, and ${\bar{\boldsymbol a}}_{av, 1}^2$ is in the third group. The minimum value $||{\bar{\boldsymbol a}}_{av, 4}^2||$ is in the first groups. Thus PDCs 1, 3, and 5 are treated as malicious. This shows that even with smaller value of $\rho$, there may be instances when S-ADMM may identify a wrong PDC.
The bottom figure of Fig. \ref{s_rr_admm_after} shows the first element of ${\boldsymbol w}_{rr,j}^k$ using RR-ADMM with $\rho = 10^{-6}$. After 6 iterations, ${\boldsymbol w}_{rr, 2}^2(1) - {\boldsymbol w}_{rr, 2}^1(1) = 10^{-10}$ and ${\boldsymbol w}_{rr, 3}^3(1) - {\boldsymbol w}_{rr, 3}^2(1) = 2\times10^{-10}$. The central PDC calculates ${\boldsymbol w}_{rr, 4}^4 - {\boldsymbol w}_{rr, 4}^3 = {\boldsymbol 0}_{2n \times 1}$, ${\boldsymbol w}_{rr, 5}^5 - {\boldsymbol w}_{rr, 5}^4 = {\boldsymbol 0}_{2n \times 1}$, and ${\boldsymbol w}_{rr, 1}^6 - {\boldsymbol w}_{rr, 1}^5 = {\boldsymbol 0}_{2n \times 1}$, which means PDCs 1, 4, and 5 are non-malicious. RR-ADMM, in this case, therefore identifies all the malicious PDCs successfully. This example shows that RR-ADMM, in general, is more robust to the value of $\rho$ than S-ADMM.

\begin{figure}[!t]
\centering
\includegraphics[width=8.5cm,height=6.5cm]{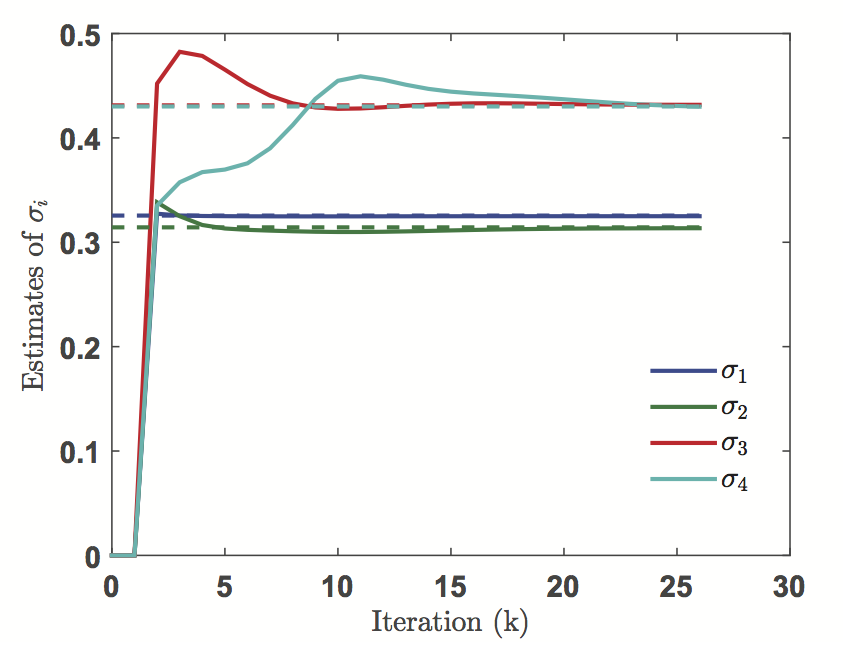}
\caption{Values of the real part of the four estimated inter-area modes before and after detection using RR-ADMM. True value of $\sigma_i$ is shown by dashed lines. }
\label{rr_admm_dec_small}
\end{figure}

Fig. \ref{rr_admm_dec_small} shows the estimation of $\sigma_i$ when the central PDC uses RR-ADMM based on the dual variable ${\boldsymbol w}^k_{rr,i}$ to catch the malicious PDCs 2 and 3. These PDCs are detected at iteration 7, onwards from which the central PDC cuts them off and continues S-ADMM with the non-malicious PDCs. The final values of the estimates of $\sigma_i$ match with their true values again.


\section{Conclusion}

In this paper we developed four algorithms for revealing the identities of malicious attackers in power system mode estimation loops. The attack is modeled in terms of data manipulations in the messages passed between distributed estimators. {The need for a distributed architecture arises primarily from the computational bottlenecks of this problem. For example, a single centralized estimator may take an enormous amount of time if it has to process hundreds of PMU measurements. Dividing the computational burden among multiple estimators soothes this complexity. The upshot, however, is that increasing the number of estimators also means increasing the number of communication links for information exchange, and therefore, increasing the risk of cyber-attacks in these channels. Moreover, in the centralized case, the entire estimation will fail if the estimator is attacked. In the distributed case, however, even if any of the estimators are attacked, the estimation can still be sustained using the detection algorithms presented in the paper. The speed of convergence may suffer due to PDCs being cut off, but the accuracy of the estimation will still be guaranteed following the consensus nature of the problem.} The specific application presented here is wide-area oscillation estimation, although the same technique can be used for other applications such as distributed dispatch, optimal power flow, and distributed control problems as well.
Future work in this area will include generalizing the detection and correction procedures to more specialized attack models such as denial-of-service \cite{alefiya}, jamming attacks on GPS signals \cite{sonia} and \cite{Lee2013}, and eavesdropping..

\appendices

\section{Expressions of $f(\cdot)$ functions in (\ref{differenceav_rr}):}
\label{appendix 2}

The derivation of the functions $f(\cdot)$ in (\ref{differenceav_rr}) follows from induction, shown in the three steps as follows.

Step 1: When $k = 1$, $f^1(A_j^1, \ldots, A_j^k,\rho) = f^2(A_j^2, \ldots, A_j^k,\rho) = \ldots = f^{k-1}(A_j^{k-1}, \ldots, A_j^k,\rho) = 0 $ and $f^k(A_j^k) = 2 A_j^1$.
Thus
\begin{small}
\begin{align}
{\boldsymbol a}_{av, j}^2  - {\boldsymbol a}_{rr, j}^2 = 2\rho A_j^1({\boldsymbol z}_{av}^1 - {\boldsymbol z}_{rr}^1).
\end{align}
\end{small}

Step 2: For all $k$, $k \ge 2$, assume
\begin{align}\label{aavarr}
&{\boldsymbol a}_{av, j}^{k} - {\boldsymbol a}_{rr, j}^{k} =\rho \sum\limits_{i = 1}^{k-1} f^i (A_j^1, \ldots, A_j^k,\rho)({\boldsymbol z}_{av}^i - {\boldsymbol z}_{rr}^i )
\end{align}

Step 3: Under the assumption of Step 2, the difference ${\boldsymbol a}_{av,j}^{k+1} - {\boldsymbol a}_{rr,j}^{k+1}$ will be proved to satisfy Eq. \eqref{differenceav_rr}.

According to \eqref{p3_2}, the differences of ${\boldsymbol a}_{rr}^{k+1} - {\boldsymbol a}_{rr}^{k}$ and
${\boldsymbol a}_{av}^{k+1} - {\boldsymbol a}_{av}^{k}$ are given by
\begin{align}
\frac{{\boldsymbol a}_{rr}^{k+1}}{A_j^k} - B_j^k - \frac{{\boldsymbol a}_{rr}^{k}}{A_j^{k-1}} + B_j^{k-1} &= \rho\left(-{\boldsymbol a}_{rr, j}^k + 2{\boldsymbol z}_{rr}^k - {\boldsymbol z}_{rr}^{k-1}\right),\\
\frac{{\boldsymbol a}_{av}^{k+1}}{A_j^k} - B_j^k - \frac{{\boldsymbol a}_{av}^{k}}{A_j^{k-1}} + B_j^{k-1} &= \rho\left(-{\boldsymbol a}_{av, j}^k + 2{\boldsymbol z}_{av}^k - {\boldsymbol z}_{av}^{k-1}\right).
\end{align}
Then the difference between ${\boldsymbol a}_{av}^{k+1}$ and ${\boldsymbol a}_{rr}^{k+1}$ is
\begin{align}\label{aavkarrk}
{\boldsymbol a}_{av}^{k+1} - {\boldsymbol a}_{rr}^{k+1} &= \rho A_j^k({\boldsymbol a}_{av}^{k} - {\boldsymbol a}_{rr}^{k})+\frac{A_j^k}{A_j^{k-1}}({\boldsymbol a}_{av}^{k} - {\boldsymbol a}_{rr}^{k}) \nonumber\\
&+ 2 \rho A_j^k({\boldsymbol z}_{av}^k - {\boldsymbol z}_{rr}^k) - A_j^k \rho ({\boldsymbol z}_{av}^{k-1} - {\boldsymbol z}_{rr}^{k-1}).
\end{align}
Substituting \eqref{aavkarrk} into \eqref{aavarr}, the difference between ${\boldsymbol a}_{av}^{k+1}$ and ${\boldsymbol a}_{rr}^{k+1}$ can be expressed as in \eqref{differenceav_rr}.

\vspace{-0.5in}

\begin{IEEEbiographynophoto}{Mang Liao}
received her BE degree in Electrical Engineering from China Civil Aviation, China in 2009 and PhD degree in Electrical Engineering from Beihang University, China in 2014. She is currently pursuing her second PhD degree in Electrical and Computer Engineering at North Carolina State University, Raleigh, NC. Her research interests are in cyber-security and intrusion detection in power systems monitoring and control.
\end{IEEEbiographynophoto}
\vspace{-0.5in}

\begin{IEEEbiographynophoto}{Aranya Chakrabortty}
received his PhD degree in Electrical Engineering from Rensselaer Polytechnic Institute, Troy, NY in 2008. He is currently an Associate Professor in the Electrical and Computer Engineering department of North Carolina State University, Raleigh, NC, where he is also affiliated to the FREEDM Systems Center. His research interests are in all branches of control theory with applications to power systems, especially in wide-area monitoring and control of large power systems using Synchrophasors. He received the NSF CAREER award in 2011.
\end{IEEEbiographynophoto}

\end{document}